\documentclass{article}

\usepackage{listings}
\usepackage{arxiv}
\usepackage{adjustbox}
\usepackage[utf8]{inputenc} 
\usepackage[T1]{fontenc}    
\usepackage{hyperref}       
\usepackage{url}            
\usepackage{booktabs}       
\usepackage{amsfonts}       
\usepackage{nicefrac}       
\usepackage{nicefrac}       
\usepackage{gensymb}        
\usepackage{microtype}      
\usepackage{lipsum}		
\usepackage{graphicx}
\usepackage{natbib}
\usepackage{doi}
\usepackage{bm}
\usepackage{rotating}
\usepackage{tikz}
\usepackage{caption,color,float}

\usepackage[bb=px]{mathalfa} 

\usepackage[varvw]{newtxmath}       
\usepackage{amsmath}
\mathchardef\mhyphen="2D 
\usepackage{booktabs}
\usepackage{multirow}
\usepackage{comment}
\usepackage{diagbox}
\usepackage{soul}
\usepackage{fancyhdr}
\usepackage{wrapfig}
\usepackage{graphicx}

\usepackage{subcaption}

\usepackage{subcaption}

\title{Degradation-Aware and Machine Learning-Driven Uncertainty Quantification in Crystal Plasticity Finite Element: Texture-Driven Plasticity in 316L Stainless Steel}

\author{
    {Dinesh ~Kumar} \\
    Department of Mechanical Engineering\\
    University of Bristol\\
    Bristol BS8 1TR, UK \\
    \texttt{dinesh.kumar@bristol.ac.uk} \\
    \And
    {Eralp ~Demir} \\
    Department of Engineering Science\\
    University of Oxford\\
    Parks Road, Oxford OX1 3PJ, UK \\
    \And
    {Julio ~Spadotto} \\
    Department of Materials\\
    University of Manchester\\
    Manchester M13 9PL, UK \\
    \And
        {Kazuma ~Kobayashi} \\
    Nuclear, Plasma \& Radiological Engineering \\
    National Center for Supercomputing Applications \\
    University of Illinois at Urbana-Champaign \\
    Urbana, IL 61801, USA \\
    \And
    {Syed Bahauddin ~Alam} \\
    Nuclear, Plasma \& Radiological Engineering \\
    National Center for Supercomputing Applications \\
    University of Illinois at Urbana-Champaign \\
    Urbana, IL 61801, USA \\
    \And
    {Brian ~Connolly} \\
    Department of Materials\\
    University of Manchester\\
    Manchester M13 9PL, UK \\
    \And
    {Ed ~Pickering} \\
    Department of Materials\\
    University of Manchester\\
    Manchester M13 9PL, UK \\
    \And
    {Paul ~Wilcox} \\
    Department of Mechanical Engineering\\
    University of Bristol\\
    Bristol BS8 1TR, UK \\
    \And
    {David ~Knowles} \\
    Department of Mechanical Engineering\\
    University of Bristol\\
    Bristol BS8 1TR, UK \\
    \And
    {Mahmoud ~Mostafavi} \\
    Department of Mechanical and Aerospace Engineering\\
    Monash University\\
    Clayton, Victoria 3800, Australia \\
}

\renewcommand{\headeright}{Preprint}
\renewcommand{\undertitle}{Preprint}
\renewcommand{\shorttitle}{\textit{Degradation-Aware and Machine Learning-Driven Uncertainty Quantification
}}

\hypersetup{
pdftitle={A template for the arxiv style},
pdfsubject={q-bio.NC, q-bio.QM},
pdfauthor={David S.~Hippocampus, Elias D.~Striatum},
pdfkeywords={First keyword, Second keyword, More},
}

\begin{document}
\maketitle

\begin{abstract}
{ \small

The mechanical properties and long-term structural reliability of crystalline materials are strongly influenced by microstructural features such as grain size, morphology, and crystallographic texture. These characteristics not only determine the initial mechanical behavior but also govern the progression of degradation mechanisms—such as strain localization, fatigue damage, and microcrack initiation under service conditions. Variability in these microstructural attributes, introduced during manufacturing or evolving through in-service degradation, leads to uncertainty in material performance. Therefore, understanding and quantifying microstructure-sensitive plastic deformation is critical for assessing degradation risk in high-value mechanical systems. This study presents a first-of-its-kind machine learning-driven framework that couples high-fidelity crystal plasticity finite element (CPFE) simulations with data-driven surrogate modeling to accelerate degradation-aware uncertainty quantification in welded structural alloys. Specifically, the impact of crystallographic texture variability in 316L stainless steel weldments, characterized via high-throughput electron backscatter diffraction (EBSD) is examined through CPFE simulations on calibrated representative volume elements (RVEs). A polynomial chaos expansion-based surrogate model is then trained to efficiently emulate the CPFE response using only 200 simulations, reducing computational cost by several orders of magnitude compared to conventional Monte Carlo analysis. The surrogate enables rapid quantification of uncertainty in stress-strain behavior and identifies texture components, such as Cube and Goss, as key drivers of degradation-relevant plastic response. This framework represents a novel integration of physics-based modeling, experimental texture data, and uncertainty-aware machine learning, offering a scalable and accurate approach for predictive degradation modeling in welded nuclear-grade materials.

}

\end{abstract}

%
{ 

\section{Introduction}

The mechanical behaviour and long-term degradation resistance of metallic service components critically depend on their microstructural features, particularly grain structure. Manufacturing processes such as casting \citep{wang2023exceptional, zhao2024effects,sharifi2018grain,hossain2024sensor,almutairi2022weight}, drawing \citep{wang2020microstructure, shan2020exceptional}, extrusion \citep{donati2013grain, mayama2011crystal}, and welding \citep{mokhtarishirazabad2023predicting, dong2022microstructure} often induce non-random grain orientation distributions, generating localized crystallographic texture \citep{sofras2022tailored, kalidindi1992crystallographic,veasna2023machine}. These microstructural anisotropies affect local plasticity and play a fundamental role in initiating degradation pathways. Dislocation pileup at high-angle grain boundaries \citep{lu2020improving,winning2005transition} and transfer across low-angle boundaries \citep{liu2021response} elevate stress concentrations, accelerating damage evolution through fatigue, strain localization, and intergranular cracking \citep{huang2025localized, ritchie2000mechanics}. In critical sectors such as nuclear and aerospace, a deep understanding of how process-induced microstructures impact degradation behaviour is essential. While advanced microscopy informs design and forensic assessment, such analyses are inherently costly, destructive, and spatially limited, potentially missing representative degradation features. Thus, they are increasingly complemented by physics-based simulations. Among these, crystal plasticity finite element (CPFE) models are indispensable for capturing grain-scale degradation precursors, yet their high computational cost becomes a barrier when exploring uncertainty due to microstructural variability. This is particularly limiting when simulating large-scale welded components or service-aged alloys, where local heterogeneity governs degradation onset. Consequently, efficient modeling frameworks that maintain physical fidelity while enabling stochastic analysis are crucial for predictive degradation assessment and lifecycle management of high-value structural systems.

The integration of crystal plasticity theory within the finite element method, known as the CPFE method, not only predicts the macroscopic stress-strain curve and texture evolution, but also resolves local deformation fields that are critical to understanding early-stage degradation phenomena such as strain localization, stress concentration, and damage nucleation \citep{miehe2014variational,bulgarevich2023crystal, ma2022virtual}. However, CPFE simulations are computationally intensive, especially when modeling a wide range of microstructures to quantify the stochastic evolution of degradation under service conditions \citep{franceschetti1999inverse, butler2018machine, liu2017materials, zhao2021grain,weber2022machine,sahu2024micro}. To alleviate this cost, homogenized methods are often used in large-scale simulations due to their computational efficiency \citep{cruzado2017modeling}. Yet, these approaches average out microscale stress fields and texture effects, limiting their ability to capture the heterogeneity that drives localized degradation and ultimately governs component failure.

Although CPFE approaches can accurately capture microstructural features and heterogeneous properties of polycrystalline materials \citep{ashraf2024investigating}, their high computational cost limits their application to large-scale degradation modeling. In particular, when assessing damage evolution driven by microstructure—such as fatigue crack initiation, creep strain localization, or stress-corrosion susceptibility—performing hundreds of CPFE simulations to resolve uncertainty becomes impractical. To address this, reduced-order or surrogate modeling techniques have emerged as powerful tools for enabling efficient degradation predictions. Recent advances in machine learning (ML) have shown remarkable promise in engineering and materials science, particularly for linking complex microstructural inputs to degradation-relevant mechanical responses. Techniques such as deep learning and artificial neural networks \citep{kobayashi2024improved,kobayashi2024deep,nascimento2023machine,raman2024data,kobayashi2024explainable,kabir2024transfer,daniell2025digital,kobayashi2024physics,shen2021generic,li2024unveiling,kobayashi2024physics}, Gaussian processes \citep{ogren2024gaussian, xiong2023gaussian,kobayashi2022practical}, polynomial chaos \citep{sudret2008global,zhou2020surrogate,kobayashi2023data}, support vector machines \citep{wang2024recent}, Bayesian inference \citep{venkatraman2022bayesian}, and recurrent neural networks \citep{liu2015predictive, zhang2019extracting, farrar2012structural} have proven effective at capturing material behavior. These data-driven models are increasingly used to accelerate the discovery of materials resistant to environmental or mechanical degradation \citep{mannodi2016machine, bock2019review, paulson2018data,veasna2023machine}. ML models have successfully learned correlations between microstructure and failure-relevant performance metrics, including stress hotspots and strain localization \citep{bock2019review, cook2000combining, roirand2024understanding}. They have also been applied to inverse problems such as microstructure reconstruction from diffraction data \citep{shen2019convolutional,cherukara2018real}. For example, Mangal et al. \citep{mangal2018applied} used random forest methods to predict grain-level stress hotspots, precursors to fatigue or creep damage. Across many studies, ML-based surrogate models have offered orders-of-magnitude speedups over direct numerical simulations \citep{capuano2019smart, bulgarevich2024stress}, supporting real-time assessment of degradation risks in structural materials \citep{ali2019application, ozturk2021uncertainty}.

To overcome the limitations of high-fidelity simulations in large-scale degradation studies, surrogate modeling has emerged as a viable alternative. However, existing studies have primarily focused on using idealized or synthetic microstructures, often neglecting realistic microstructural heterogeneity derived from actual weldments. In contrast, the present work introduces a novel framework that integrates experimental texture data from electron beam welded 316L stainless steel with crystal plasticity simulations and data-driven surrogate modeling to enable degradation-aware uncertainty quantification. This study is among the first to employ high-throughput EBSD-informed statistical texture bounds as direct inputs for CPFE modeling and to construct a physics-informed surrogate using polynomial chaos expansions (PCE) tailored for texture-driven plasticity.

The surrogate model is designed not only to reproduce full-field CPFE outputs but also to identify critical texture components influencing early-stage degradation mechanisms such as strain localization and stress amplification. The novelty lies in the tight coupling between real microstructural data, mechanistic simulation, and uncertainty-aware surrogate learning, which together allow for the efficient prediction of degradation-relevant responses across a spectrum of realistic microstructural states. This approach represents a significant advance over conventional Monte Carlo or purely data-driven techniques, both in terms of computational efficiency and physical interpretability. By reducing the cost of stochastic CPFE analysis by several orders of magnitude, this method enables scalable exploration of microstructure-sensitive degradation behavior in nuclear-grade materials.

This work is part of a collaborative UKRI-funded project called SINDRI (Synergistic Utilisation of Informatics and Data-Centric Integrity Engineering), which aims to advance predictive digital technologies for assessing degradation in nuclear plant components. The study investigates uncertainty in the degradation-relevant mechanical response of 316L stainless steel by integrating CPFE modeling with machine learning-based surrogate models. We examine mesh refinement and grain count effects within representative volume elements (RVEs), evaluating the trade-off between computational efficiency and accuracy in capturing stress and strain localization. Using a statistically converged RVE, we develop a polynomial chaos-based surrogate to model the influence of texture on material response. The approach enables efficient uncertainty and sensitivity analyses of degradation precursors.

\section{Mathematical Modeling}
The model in this study includes two concepts: the CPFE method used for micromechanical simulations, and the polynomial chaos approach for surrogate modeling, both are described in detail in the following sections. 

\subsection{Crystal Plasticity Finite Element}
Finite element method solves the weak form of local equilibrium, is defined as the sum of the divergence of stress, $\boldsymbol{\sigma}$, with the body force per volume  $\bm{b}$ for a quasi-static problem:
\begin{equation}
\nabla \, \cdot \, \boldsymbol{\sigma} \, + \,\bm{b} \, = \, \bm{0}.
\label{eq1} 
\end{equation}
Multiplication of the equilibrium equation with an arbitrary weight function, $\bm{N}$, and integration over the domain followed by the use of divergence theorem calculates the force residual, $\bm{r}$, which is the difference between the internal force, $\bm{f}_{int}$, and external force, $\bm{f}_{ext}$:
\begin{equation}
\bm{r} \,=\, \bm{f}_{ext} \,-\, \bm{f}_{int}.
\label{eq2}  
\end{equation}
The weak form of external force vector contains a term related to the body force, $\bm{b}$, and another term related with surface traction, $\bm{t}$, with corresponding integration weights of $\bm{N}$ and $\bm{N}_s$, respectively:
\begin{equation}
\bm{f}_{ext}\,=\, \int_V \bm{N}^T \, \bm{ b}\,  \,\text{d}V \,+\, \int_S \bm{N}_s^T \, \bm{t}  \,\text{d}S,
\label{eq3}  
\end{equation}
and the internal force vector is computed using strain-displacement, $\bm{B}$, and stress as:
\begin{equation}
\bm{f}_{int}\,=\, \int_V \bm{B}^T \, \sigma  \,\text{d}V.
\label{eq4}
\end{equation}
In an implicit solution scheme, the residual is minimized iteratively (often using Newton-Raphson method) to find equilibrium displacements according to:
\begin{equation}
\bm{u}_{k+1}\,=\,  \bm{u}_{k} \,+\, \bm{K_T}^{-1} \, \bm{r},
\label{eq5} 
\end{equation}
in which $\bm{K_T}$ is the tangent stiffness given by:
\begin{equation}
\bm{K_T}\,=\, \int_V \bm{B}^T \, \dfrac{\partial \Delta \boldsymbol\sigma}{\partial \Delta \boldsymbol\varepsilon}  \, \bm{B} \,\text{d}V.
\label{eq6}
\end{equation}
The equilibrium is solved within an iterative solver for non-linear material properties to reveal the displacement field and related quantities such as total strain increment, $\Delta\varepsilon$, and total deformation gradient, $\bm{F}$ which the finite element code gives these quantities for small strain or large strain cases, respectively. Therefore within the user subroutine (UMAT), $\Delta\varepsilon$ and $\bm{F}$ are available and stress, $\boldsymbol{\sigma}$, and material tangent, $\frac{\partial \Delta \boldsymbol\sigma}{\partial \Delta \boldsymbol\varepsilon}$, are needed to calculate the internal force vector in Eq. \eqref{eq4} and tangent stiffness in Eq. \eqref{eq6}, respectively. Most FE codes \textsuperscript{\textregistered} use co-rotational Cauchy stress as the stress measure at the deformed configuration and engineering strains (shear strains doubled). In Voigt notation with the vectorised form of stress:
\begin{equation}
\boldsymbol\sigma^T\,=\, \left\{ \begin{array}{cccccc}
  \sigma_{11}   & \sigma_{22} & \sigma_{33} & \sigma_{12} & \sigma_{13} & \sigma_{23}  
\end{array}  \right\}.
\label{eq7}
\end{equation}
Crystal plasticity splits the total deformation, $\bm{F}$ to elastic, $\bm{F_e}$, and plastic, $ \bm{F_p}$ parts, using the multiplicative deformation for a general large deformation case assuming that the plastic deformation does not lead to volumetric change:
\begin{equation}
\bm{F} \,=\, \bm{F_e} \, \bm{F_p}, \;\; \text{with} \;\;\;\;\; \text{det}(\bm{F_p})=1.
\label{eq8}
\end{equation}
Plastic slip, $\dot{\gamma}^a$, is transformed to crystal reference coordinate system and summed over the slip systems, $a$, to give the plastic part of the velocity gradient, $\bm{L}_p$:
\begin{equation}
\bm{L}_p \,=\, \sum_a  \, \bm{s_0}^a \,\otimes \, \bm{n_0}^a \, \dot{\gamma}^a
\label{eq9}
\end{equation}
Slip direction, $\bm{s_0}^a$, and slip plane normal, $\bm{n_0}^a$ are computed at the intermediate configuration and they do not need updating with plastic deformation. In this study, the crystal plasticity problem is solved at the plastically deformed reference\cite{kalidindi1992crystallographic}, so called intermediate configuration giving:
\begin{equation}
\bm{s}^a \, = \, \bm{F_e} \, \bm{s_0}^a, \;\; \text{and} \;\;\;\;\; \bm{n}^a \, =\, \bm{F_e}^{-T} \, \bm{n_0}^a.
\label{eq10}
\end{equation}
Green-Lagrange strain, $\bm{E_e} $, is used to find the elastic strains using Eq. \eqref{eq11} in which $\bm{I}$ is the second rank identity tensor:
\begin{equation}
\bm{E_e} \, = \, \dfrac{\bm{F_e}^T \, \bm{F_e} \,-\, \bm{I}}{2}  
\label{eq11}
\end{equation}
The solution is based on the Second Piola-Kirchoff (PK2) stress, $\bm{S}$, that is obtained using $\bm{C_0}$ fourth rank elasticity tensor at the intermediate configuration\footnote{Inner product or double contraction operation is indicated with the symbol $:$.}:
\begin{equation}
\bm{S} \, = \, \bm{C_0} \,:\, \bm{E_e}   
\label{eq12}
\end{equation}
PK2 stress at the intermediate configuration is related to the Cauchy stress at the deformed configuration through appropriate transformations as:
\begin{equation}
\bm{\sigma} \, = \, \bm{F_e} \,\bm{S}\, \bm{F_e}^{T}  \, / \, \text{det}(\bm{F_e})
\label{eq12}
\end{equation}
The resolved shear stress (RSS), $\tau^a$, that is the work conjugate driving-force for slip \cite{kalidindi1992crystallographic} is the projection of stress on to the slip systems:
\begin{equation}
\tau^a \, = \,  \bm{F_e}^T  \, \bm{F_e} \, \bm{S} \,:\, \bm{s_0}^a \otimes  \bm{n_0}^a \,\approx\, \bm{s_0}^a \, \cdot \, \bm{S}  \, \cdot \, \bm{n_0}^a
\label{eq13}
\end{equation}
The slip rates are governed by power law (as it facilitates convergence in comparison with exponential function) in which $\tau_c^a$ represents the critically resolved shear stress (CRSS) and $X^a$ denotes the back-stress as the state variables. We have included the back-stress in the formation as the ultimate goal of the work is estimating the uncertainty in the cyclic response of the material which has a combined isotropic and kinematic hardening behaviour. $\dot\gamma_0$ and $n$ represent the reference slip rate and power law exponent:
\begin{equation}
\dot\gamma^a \,=\, \dot\gamma_0 \,\left( \left|  \tau^a \,-\, X^a \right| \,/ \, \tau_c^a \right)^n \, \text{sgn}(\tau^a)
\label{eq14}
\end{equation}
CRSS is linked with the state variable statistically-stored dislocation density (SSD), $\varrho_{SSD}^a$, using Taylor equation in which $\alpha$, $G$, $b$, $D$, and $L^a$ denoting the geometric factor, shear modulus, Burgers vector, grain size factor, and mean slip distance in a grain, respectively:
\begin{equation}
\tau_c^a \,=\, A \, G \, b \, \left(\sqrt{\varrho_{SSD}^a} \,+\,  \dfrac{B}{L^a} \right)
\label{eq15}
\end{equation}
The grain size dependence is introduced to the evolution of SSD density with the additional term, $L^a$ to recreate dislocation pileup and recreate a Hall-Petch effect. $L^a$ is computed by fitting ellipsoids to each grain and projection of the principal axes of the ellipsoid for each slip system using a special projection operator as described in detail elsewhere \cite{demir2023grain}. Kocks-Mecking-Estrin hardening model \cite{mecking1981kinetics, estrin1984unified} governs the evolution of SSDs in which $C$ and $D$ are hardening and softening model constants respectively:
\begin{equation}
\dot{\varrho}_{SSD}^a \,=\, \left[\dfrac{C}{b}  \,  \left(\sqrt{\varrho_{SSD}^a} \,+\,  \dfrac{B} {L^a}  \right)  \,-\, D \varrho_{SSD}^a \right] \,  | \dot{\gamma^a} |
\label{eq16}
\end{equation}
The other state variable backs-stress, $X^a$, evolves according to Armstrong-Frederic (AF)  model \cite{armstrong1966mathematical} in which $E$ and $F$ represent kinematic hardening and softening constants:
\begin{equation}
 X^a \,=\, E \, \dot{\gamma^a} \, -  \, F \,X^a\,   | \dot{\gamma^a} |
\label{eq17}
\end{equation}
The details of the solution of crystal plasticity and an analytical expression for the material tangent is provided within the reference \cite{demir2023grain}.

\subsection{RVE and Simulation Process}

The initial step is to generate a representative volume element (RVE) that mimics key microstructural features identified through experimental characterization, particularly those known to influence early-stage degradation mechanisms. The RVE contains material grains matched to the experimentally observed grain size distribution; Figure \ref{fig:RVE} shows an example RVE in which each grain is depicted with a different color. In computational simulations, discretizing the domain into a mesh of finite elements enables numerical solution of governing equations such as those arising in structural mechanics. The resolution of this mesh is particularly important in degradation-aware modeling, as it governs the ability to capture localized deformation fields associated with damage precursors like strain localization and stress concentration. Therefore, mesh refinement is performed as the first step of this study.

Next, the CPFE model—known to have numerous sensitive parameters—is calibrated against reference data to ensure predictive accuracy. This step is critical for degradation-focused simulations, where miscalibration can obscure damage-relevant phenomena. A subsequent analysis evaluates the influence of grain count and RVE size on macromechanical response, as previous studies \citep{knezevic2014three, yaghoobi2021crystal} have shown that RVE fidelity strongly affects the capture of degradation-related strain heterogeneities. The grain number required for convergence is strain-dependent \citep{chen2022multiscale, lim2019investigating}, which is particularly relevant when simulating components under varying service loads. Engineering stress is computed from surface reaction forces, while engineering strain is calculated from applied displacements. The RVE that yields a statistically converged mechanical response is then used to investigate how crystallographic texture drives degradation-relevant mechanical variability. Results are presented in the following sections.

\begin{figure}[htbp]
\centering
\includegraphics[width=0.5\textwidth]{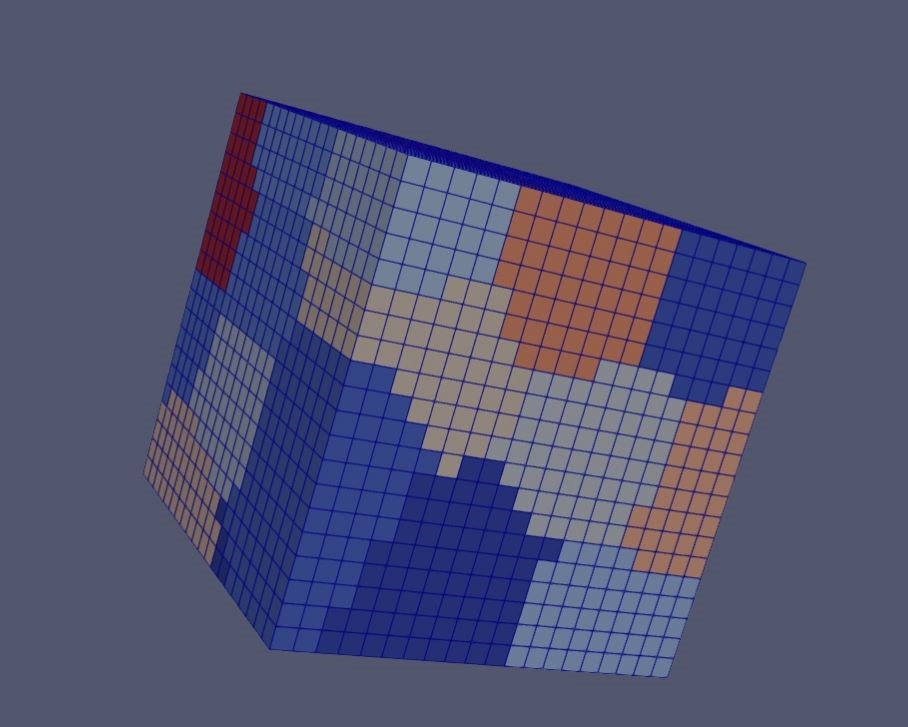}
\caption{A representative volume element with grains for CPFE simulations}
\label{fig:RVE}
\end{figure}

\subsection{Polynomial Chaos Approach for Surrogate Modeling}

Surrogate modeling is a highly impactful technique in computational science and engineering for approximating complex, computationally intensive models \cite{alizadeh2020managing, forrester2008engineering, qian2006building}. Among the various approaches, the polynomial chaos (PC) method \cite{wiener_chaos_1938, xiu_2002} is widely recognized for its simplicity in constructing surrogate models that can efficiently predict system behavior under uncertainty. This approach, grounded in polynomial chaos theory, is particularly suited for uncertainty quantification, sensitivity analysis, and reliability assessment in degradation-aware simulations \cite{kumar2021quantitative, kumar2022multi,kobayashi.2023,Blatman2008Sparse, kumar2016efficient, kumar2020efficient,kumar.2023}.

Polynomial chaos involves building a functional approximation of a model’s response using orthogonal polynomial basis functions. The method has broad utility in engineering, including areas such as fluid dynamics, nuclear design, and composite material behavior, where degradation mechanisms often arise from microstructural uncertainty \cite{kumar2021quantitative, kumar2016efficient, kumar2020efficient, najm2009uncertainty}. In the PC framework, input random variables representing microstructural heterogeneity—such as texture—are expanded using orthogonal polynomials to generate a Polynomial Chaos Expansion (PCE). This expansion, akin to a Fourier series but applied in probability space, enhances the robustness of predictions and captures variability in degradation-relevant responses \cite{kumar2016efficient, najm2009uncertainty}.

In the context of a model with $n$ input parameters and a response $u(x)$, the system response under uncertainty is approximated as:

\begin{eqnarray}\label{eq:pc}
u(\pmb \xi) = \sum_{i=0}^N u_i\psi_i(\pmb \xi)
\end{eqnarray}

Here, $u_i$ are deterministic coefficients, and $\psi_i$ are multidimensional orthogonal polynomials based on random variables $\pmb \xi = \left\{\xi_1,\xi_2, \ldots, \xi_n \right\}$. These random inputs represent statistical variability in microstructural features that affect degradation processes.

In meta-modeling with polynomial chaos, polynomial coefficients are computed using numerical methods such as projection, regression, or collocation \cite{kumar2016efficient, najm2009uncertainty,foutch2025ai,kobayashi2024ai}. The process begins by sampling the input space, generating a fixed number of realizations of the uncertain parameters (e.g., texture components). Based on the sample set, model outputs are computed and used to form the system of equations for coefficient estimation. The resulting surrogate captures how microstructural variation propagates into stress or strain variability—critical indicators of early-stage degradation. The regression method is commonly employed to establish this mapping between inputs and model response, offering a detailed description of degradation-relevant variability.

Let $M$ be the number of samples generated for $\pmb\xi$ (i.e., $\xi^1, \xi^2, \ldots, \xi^M$), and $u(x;\pmb \xi^i)$ the corresponding model response for each input. Using Equation \ref{eq:pc}, we write:

\begin{eqnarray}\label{eq:pcreg}
u(x;\pmb \xi^1) = u_0\psi_0(\pmb \xi^1)+ u_1(x)\psi_1(\pmb \xi^1)+ u_2(x)\psi_2(\pmb \xi^1)+....+ u_N(x)\psi_N(\pmb \xi^1)\\
u(x;\pmb \xi^2) = u_0\psi_0(\pmb \xi^2)+ u_1(x)\psi_1(\pmb \xi^2)+ u_2(x)\psi_2(\pmb \xi^2)+....+ u_N(x)\psi_N(\pmb \xi^2)  \nonumber\\
----------------- \nonumber\\
-----------------\nonumber\\
-----------------\nonumber\\
----------------- \nonumber\\
u(x;\pmb \xi^M) = u_0\psi_0(\pmb \xi^M)+ u_1(x)\psi_1(\pmb \xi^M)+ u_2(x)\psi_2(\pmb \xi^M)+....+ u_N(x)\psi_N(\pmb \xi^M)  \nonumber
\end{eqnarray}

The above system is expressed in matrix form as:

\begin{eqnarray}
& \left( \begin{array}{cccccc}
\psi_0({\pmb \xi}^1)&\psi_1({\pmb \xi}^1)&\psi_2({\pmb \xi}^1)& .& .&\psi_P({\pmb \xi}^1)\\
\psi_0({\pmb \xi}^2)&\psi_1({\pmb \xi}^2)&\psi_2({\pmb \xi}^2) &.& .&\psi_P({\pmb \xi}^2)\\
.&.&.& . &.&.\\
.&.&.& . &.&.\\
.&.&.& .& .&.\\
\psi_0({\pmb \xi}^{M})&\psi_1({\pmb \xi}^{M})&\psi_2({\pmb \xi}^{M}) &.& .&\psi_P({\pmb \xi}^{M})\end{array} \right)
\left( \begin{array}{c}
{u}_0\\
{u}_1\\
.\\
.\\
.\\
{u}_P\end{array} \right)
=
\left( \begin{array}{c}
u({\pmb \xi}^1)\\
u({\pmb \xi}^2)\\
.\\
.\\
.\\
u({\pmb \xi}^{M})\end{array} \right)
\end{eqnarray}
or
\begin{equation} \label{eq:21}
\pmb D U =\pmb B
\end{equation}\\

When $M > P$, the least-squares regression approach is used to solve the system:

\begin{equation}
U=(\pmb D^T \pmb D)^{-1}\pmb D^T \pmb B
\end{equation}\\

Once constructed, the polynomial chaos surrogate enables rapid estimation of statistical moments (mean, variance) of the output, such as stress or strain. These are directly linked to degradation indicators. Additionally, Sobol indices derived from the surrogate provide insight into which uncertain inputs—e.g., texture modes—most influence degradation-relevant response variability.

\section{Experiments}
The material used in this study was a commercial 316L austenitic stainless steel with the following chemical composition (wt.\%): 0.016C-16.56Cr-1.27Mn-2.04Mo-0.04N-10.04Ni-0.031P-0.002S-0.37Si-bal.Fe. To ensure weld quality and minimize structural discontinuities that could influence degradation initiation, the material was welded in the horizontal welding position (PC or 2G) using an electron beam welding process. Prior to welding, the plates were cleaned to remove any contaminants. Welding was performed under vacuum at pressures of $6.0 \times 10^{-3}$ mbar or lower. The beam operated at 80 kV, with a current of 160 mA and a speed of 4.16 mm/s.  

For EBSD analysis, the sample was metallographically prepared, with a final polishing step performed using a 0.04 µm colloidal silica suspension (OPS) in a vibropolisher (Vibromet 2 – Buehler). The EBSD analysis was conducted using a scanning electron microscopy (SEM) Thermo Scientific APREO C equipped with Symmetry 2 EBSD detector (Oxford Instruments), operating at 20 kV, with a beam current of 51 nA, a step size of 2.5 µm, and the sample tilted at 70o related to the electron beam. Data acquisition and analysis were performed using Aztec and Aztec Crystal softwares (Oxford Instruments), respectively.


The large EBSD map in Figure 1a presents the inverse pole figure (IPF) in the Z-direction of a cross-section from a 316L austenitic stainless steel plate, welded using the electron beam (EB) welding process. This map is composed of 952 individual EBSD maps, stitched together to cover the entire 25 mm thickness of the 316L plate. The map reveals the characteristic microstructure generated by the EB welding process. The fusion zone exhibits elongated columnar grains aligned with the solidification direction, driven by the thermal gradient. These grains are critical as they influence local mechanical response and may promote degradation via stress concentration or crack propagation paths.

Mesoscale EBSD imaging enables a detailed analysis of microstructural evolution across welded regions, particularly where degradation is prone to initiate due to textural anisotropy. An example of this is presented in Figure 2b, where an IPF-Z map, extracted from the EBSD dataset in Figure 2a, highlights the top of the welded region. As shown in Figure 2b, the resulting microstructure reveals different regions based on grain characteristics: (1) equiaxed face-centered cubic (FCC) grains in the parent metal (PM), (2) large FCC grains in the heat-affected zone (HAZ), (3) smaller FCC equiaxed grains at the transition between the HAZ and fusion zone (FZ), where lots of grains would have nucleated at the start of solidification, (4) columnar FCC grains resulting from the solidification of the weld pool, and (5) the center line of the welded region. The two areas outlined by the black boxes in Figure 2b correspond to the regions displayed in Figure 2. Figures 2c and 2e show the pole figures (PFs) for the parent metal and weld metal, respectively, while Figures 1d and 1f display the inverse pole figures (IPFs) for the parent metal and weld metal. In the case of the parent metal, both the PFs and IPFs indicate random crystallographic orientations of the grains, suggesting the absence of a strong texture. In contrast, the results for the weld metal reveal a more pronounced texture, indicating that the grains preferentially aligned in specific crystallographic orientations during solidification, particularly along the {100} direction, conditions known to influence strain localization and early-stage degradation.

\begin{figure} 
    \centering

    \begin{subfigure}{0.9\textwidth}
        \centering
        \includegraphics[width=\textwidth]{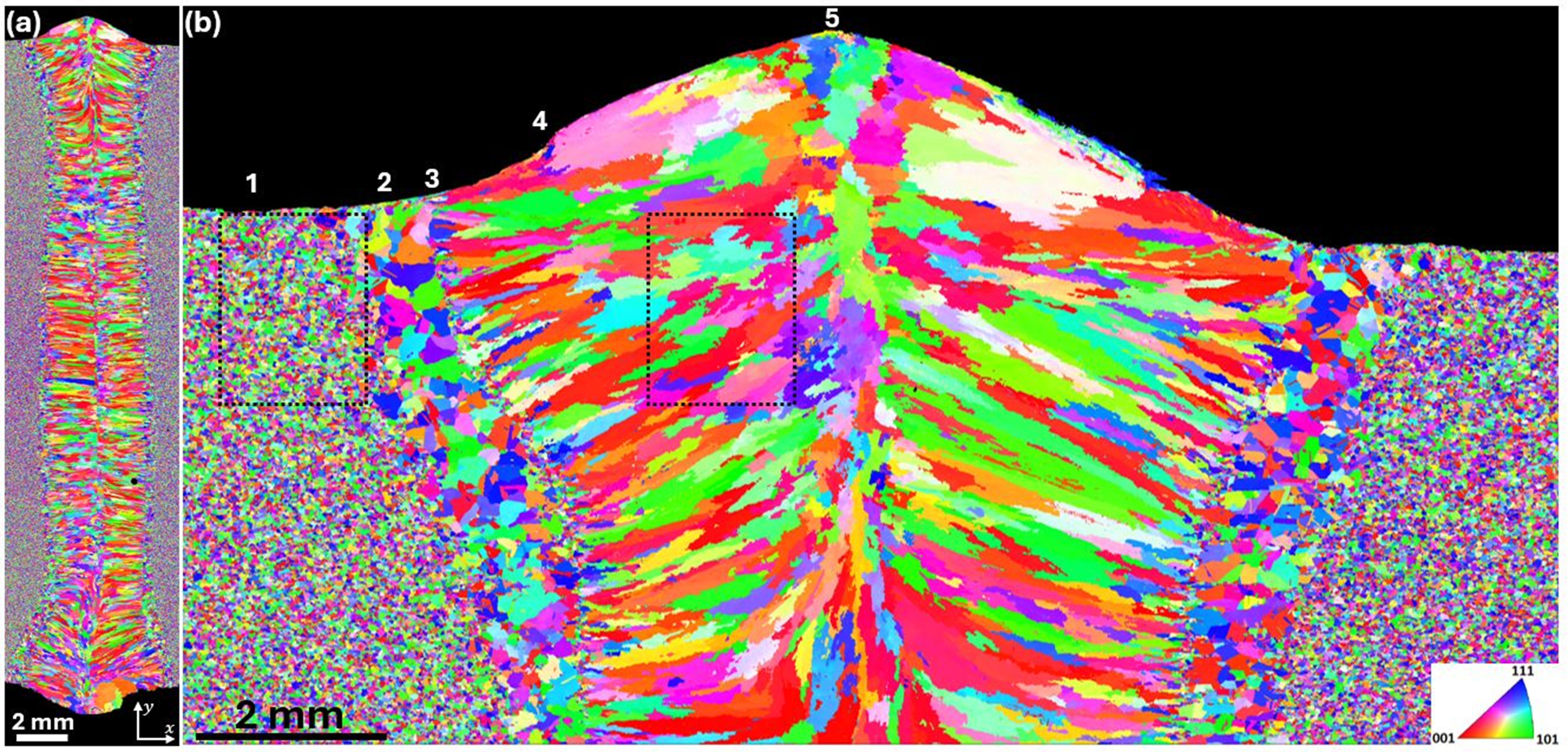}
        \end{subfigure}

    \vspace{2pt} 

    \begin{subfigure}{0.65\textwidth}
    \centering
    \includegraphics[width=\textwidth]{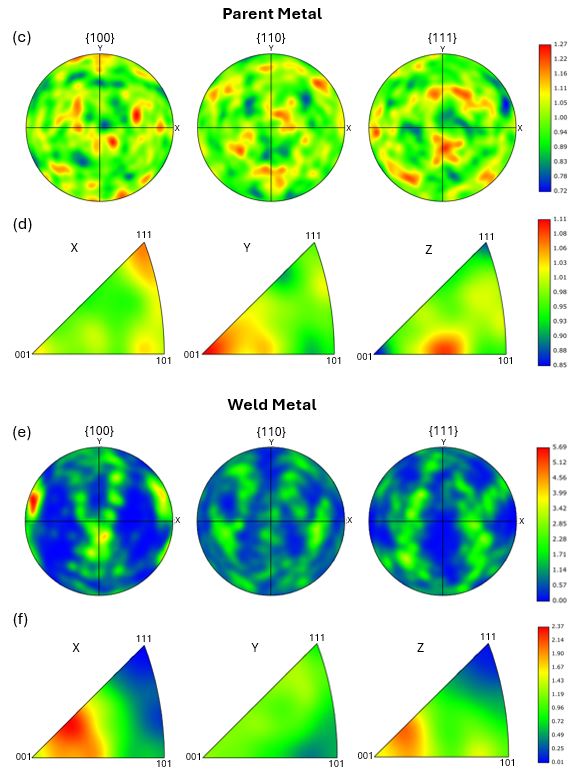}    
    \end{subfigure}

    \vspace{-2pt} 
    \caption{Large EBSD maps (IPF-Z) of 316L austenitic stainless steel welded by electron beam process. (a) Entire cross-section (25 mm thickness), (b) top region from (a), (c) pole figures and (d) inverse pole figures of parent metal, (e) pole figures and (f) inverse pole figures of weld metal.}
\label{fig:EBSDMaps}    
\end{figure}

\section{Results and Discussion}

\subsection{Mesh Convergence}

As in any mathematical modeling in which the governing equations are discretised before solution, finite element method requires a convergence study with respect to mesh refinement as a first step. This would make sure that there is acceptable level of error in the model responses associated with mesh convergence. In the context of degradation-aware simulations, mesh refinement is especially important as local stress and strain gradients—often precursors to damage—can only be captured with sufficient resolution. A finer mesh, composed of more elements, typically leads to improved accuracy due to better geometric and physical representation of the system \cite{kita2001error, liu2015adaptive, hughes2005isogeometric}. However, this comes at the cost of increased computational load.

This study involves the use of RVEs of two different sizes and grains. For the  case presented in Figure \ref{fig:MeshReine1}, an RVE of size 40 $\mu$m and 25 grains is used for mesh refinement. This number of grains is thought not to be enough for representing the macromechanical behaviour based on previous studies \citep{knezevic2014three, yaghoobi2021crystal}. It provides a suitable basis for assessing how mesh resolution influences the ability to capture degradation-relevant stress fields. Four combinations of element sizes and element counts were tested. To achieve this, we have explored four combinations of element sizes and the number of elements, which have been systematically tabulated in Table \ref{tab:meshRefine40} along with the number of elements per grain. Figure \ref{fig:MeshReine1} also provides a comprehensive presentation of stress-strain curves for all four cases, with an expanded view of the elastic region presented on the right side of the figure. Subsequently, we replicated the mesh refinement and convergence study process for another RVE of size 80 $\mu m$, which contains approximately 190 grains. The detailed results of this mesh refinement process are depicted in Figure \ref{fig:Meshrefine2} and are also presented in Table \ref{tab:meshRefine80}. Our findings from both cases strongly indicate that an RVE with a resolution of approximately 40 elements per grain ensures convergent in-grain stress distributions from crystal plasticity simulations. This finding is consistent with the reference \cite{lim2019investigating} for strain levels less than 5\%, and support the mesh design used in this study for degradation-focused crystal plasticity simulations.

\begin{figure}[htbp]
\centering
\includegraphics[width=0.9\textwidth]{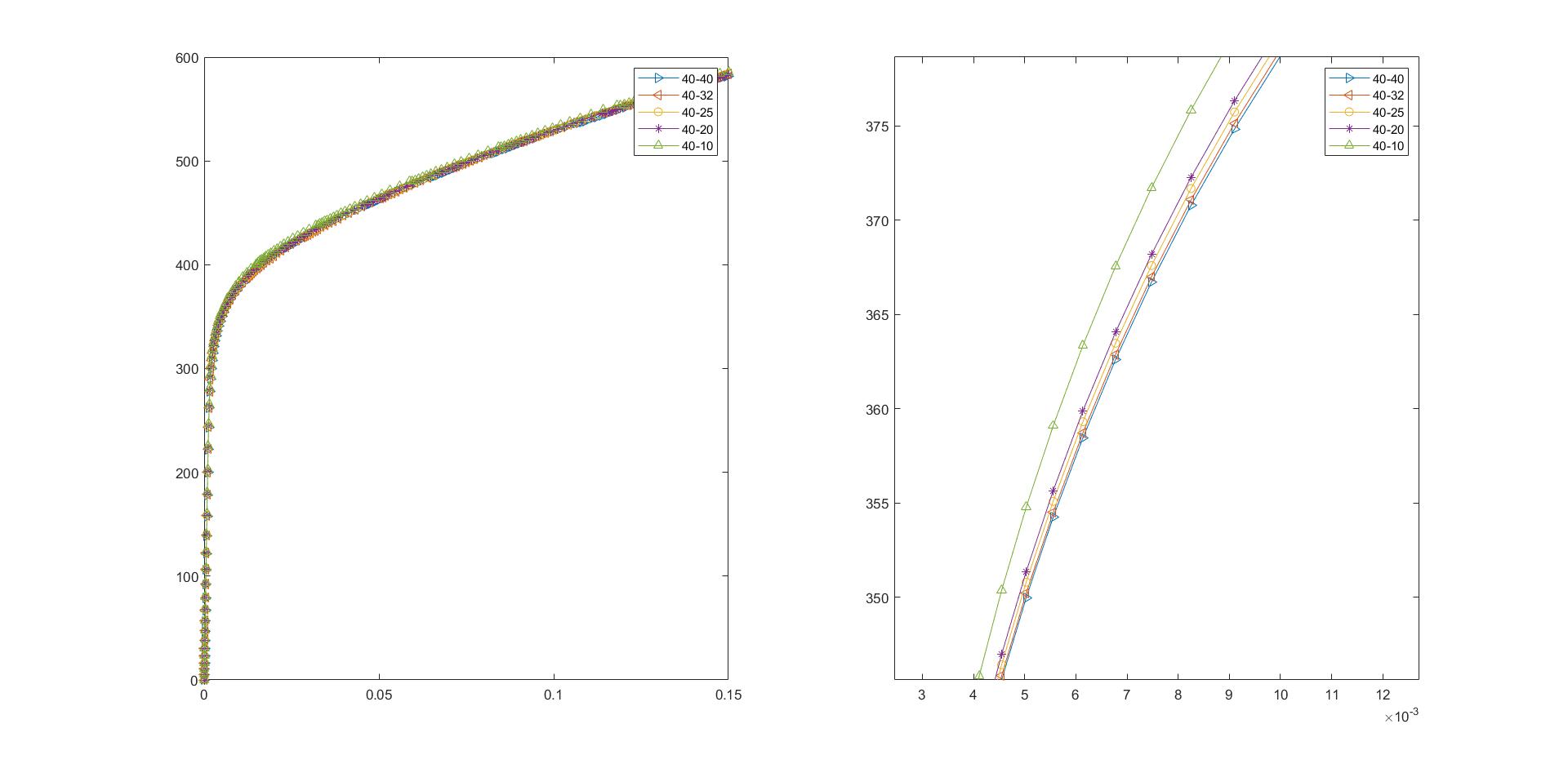}
\caption{Mesh refinement: RVE size of 40 $\mu$m that contains 25 grains (average grain size of 20 $\pm$ 8 $\mu$m)}
\label{fig:MeshReine1}
\end{figure}

\begin{figure}[htbp]
\centering
\includegraphics[width=0.9\textwidth]{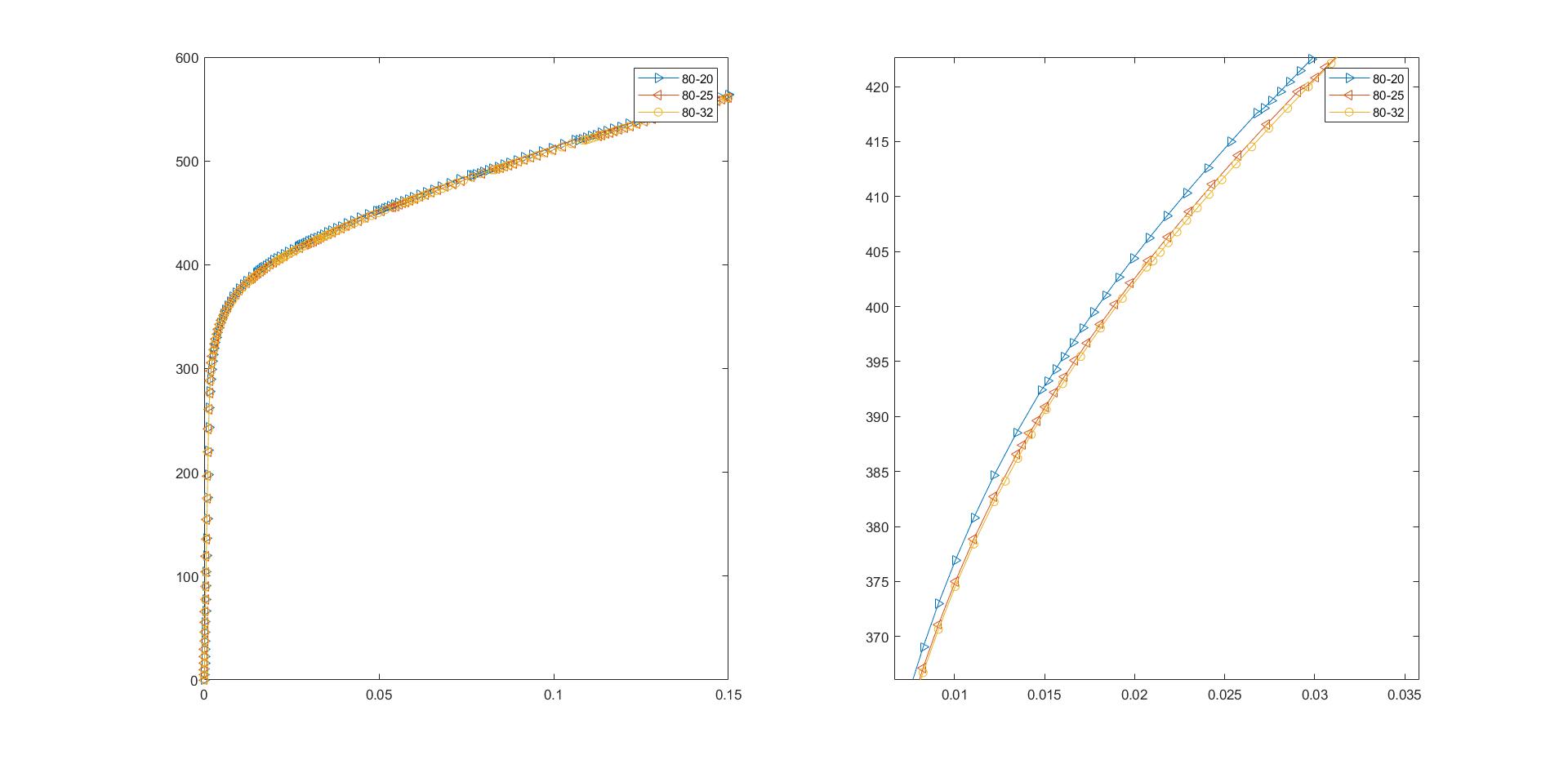}
\caption{Mesh refinement: RVE size of 80 $\mu$m that contains 25 grains (average grain size of 20 $\pm$ 8 $\mu$m)}
\label{fig:Meshrefine2}
\end{figure}

\begin{table}[!ht]
\centering
\caption{Mesh refinement for RVE size of 40 $\mu$m}
\begin{tabular}{lllllll}
 \hline\\
Elements(e\textsuperscript{3}) & Element size ($\mu$m), n=L/e & Elements per grain\\
     \hline\\
10\textsuperscript{3} & 4 & 40 \\
20\textsuperscript{3} & 2 & 320 \\
25\textsuperscript{3} & 1.6 & 625 \\
32\textsuperscript{3} & 1.25 & 1310 \\
\hline\\
\end{tabular}
\label{tab:meshRefine40}
\end{table}

\begin{table}[!ht]
\centering
\caption{Mesh refinement for RVE size of 80 $\mu$m}
\begin{tabular}{lllllll}
 \hline\\
Elements(e\textsuperscript{3}) & Element size ($\mu$m), n=L/e & Elements per grain\\
     \hline\\
20\textsuperscript{3} & 4 & 42 \\
25\textsuperscript{3} & 3.2 & 82 \\
32\textsuperscript{3} & 3.2 & 172 \\
\hline\\
\end{tabular}
\label{tab:meshRefine80}
\end{table}

\subsection{Model Calibration}

The CPFE model utilized in the analysis comprises a large number of parameters, with each playing a role in the accurate calibration and effective use of the model. One would expect all parameters in a model to be identified experimentally or theoretically calculable. However, many parameters in CPFE constitutive laws are difficult to be extracted (e.g. geometrically necessary dislocation density per slip system from high angular resolution electron back-scatter diffraction \citep{horton2023inclusion, britton2012stress}) or can be measured using various electron microscopy techniques (e.g. statistically stored dislocation density) \cite{muransky2019measurement, gallet2023experimental}. With individual experiments featuring different parameter values, their impact on CPFE outputs can vary significantly. The visual representation in Figure \ref{fig:CPFE_50} showcases 50 CPFE simulations with diverse model parameters. In degradation-aware simulations, parameter sensitivity becomes even more critical, as inaccurate calibration may obscure local plasticity behavior—such as strain localization or stress amplification—that drive early-stage damage.

Calibrating CPFE models are, therefore, a numerically undetermined problem with a nonunique solution. Hence, meticulous adjustment of parameters is essential to ensure alignment between the simulation output and specific experimental data aided by theoretical and physical understanding of the parameters to ensure they are within degradation-relevant ranges and not treated solely as empirical fitting constants. Thus, the iterative determination of parameter values, drawing from expert insights, or their calibration using advanced numerical optimization techniques, forms the basis of this process \cite{hill2000methods, beven1992future, deka2006crystal,veasna2023machine}. This becomes particularly important when using the calibrated model to evaluate degradation precursors under service-like loading conditions.

The automation of all simulation tools is vital for seamless calibration. By considering five key model parameters, extracting stress-strain data from the CPFE simulation, and computing the objective function as the root mean square error between experimental data and CPFE results, we developed an optimization process. This entails repetitive iteration of the objective function with respect to the given model parameters, aiming to minimize the root mean square error. The subsequent extraction of calibrated parameters effectively minimizes the difference between experimental results and CPFE simulation outputs. Figure \ref{fig:CPFE_Calibration} visually traces the optimization of the objective function (RMS error) and the resulting calibrated model parameters. Following this meticulous calibration, the model parameters are applied to CPFE simulation. The comparison is shown in Figure \ref{fig:CPFE_Calibration1} demonstrates that utilizing the calibrated model parameters yields CPFE simulation results that closely align with experimental data. This close agreement provides confidence that the CPFE model, once calibrated, can faithfully simulate mechanical responses relevant to degradation assessment.

It should be noted that the constitutive law of our CPFE model has been carefully designed so that only macro-mechanical experiments are required for its calibration. This will not be the case for other models, such as strain gradient models \citep{dorward2023calibration}, which require knowledge of the micromechanical behaviour of the material. In addition, since the experiment against which the current calibration is carried out is not cyclic, the isotropic hardening denoted in the hardening law by $\tau_c^a$ and kinematic hardening denoted by $X$ cannot be separated.

\begin{figure}[htbp]
\centering
\includegraphics[width=0.9\textwidth]{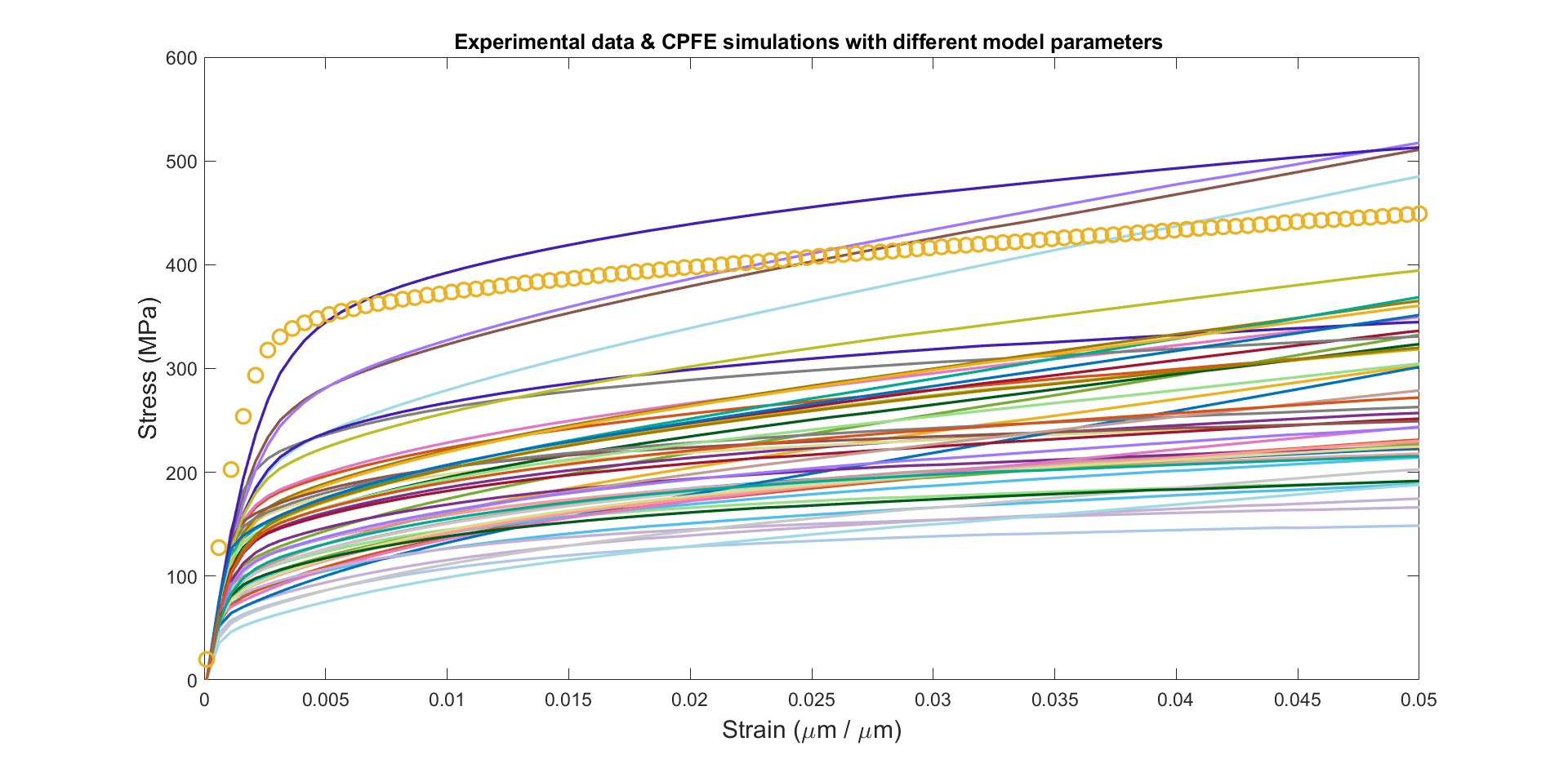}
\caption{Experimental data and CFPFE simulations with different model parameters}
\label{fig:CPFE_50}
\end{figure}

\begin{figure}[htbp]
\centering
\includegraphics[width=0.9\textwidth]{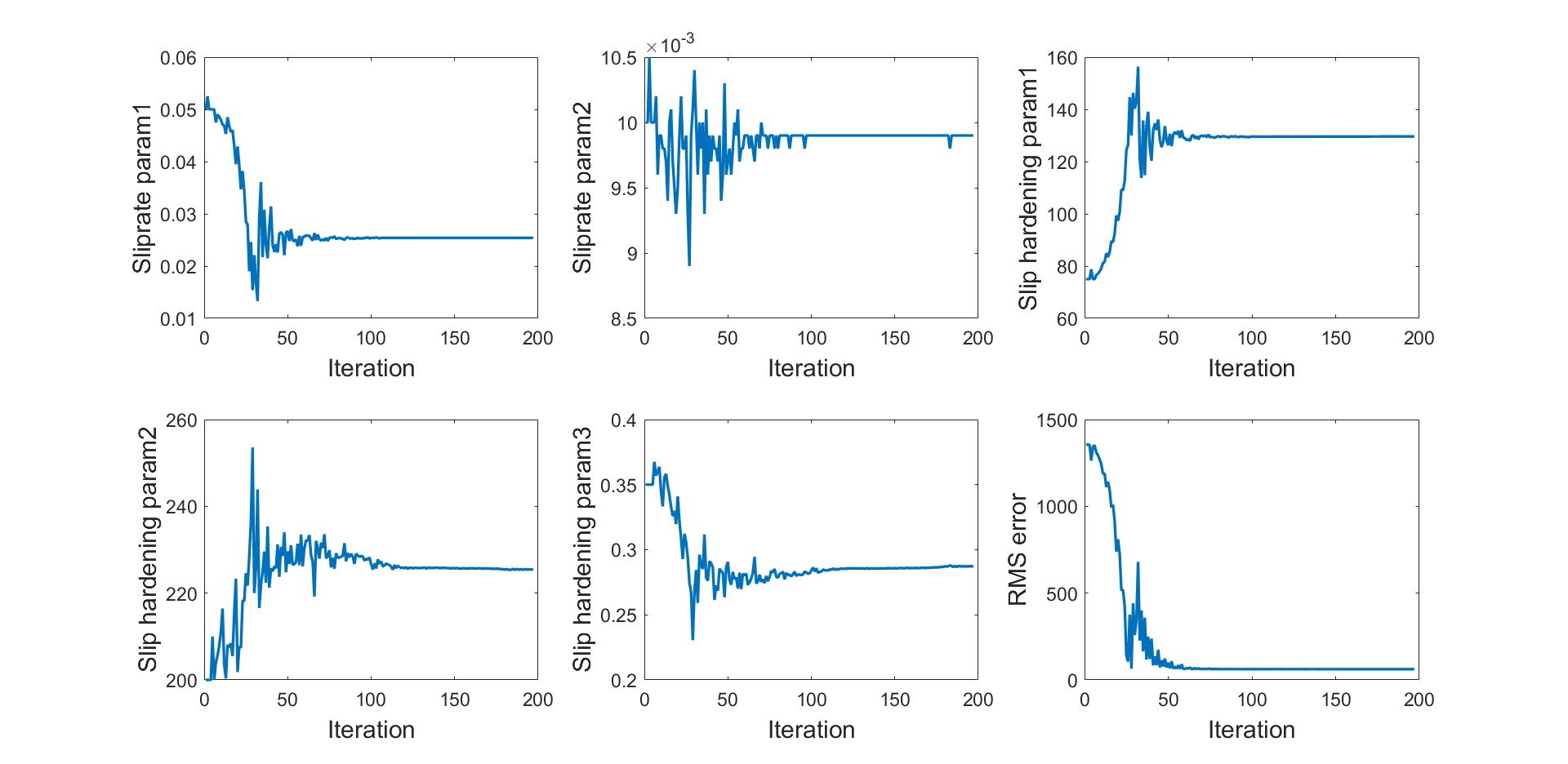}
\caption{CPFE Calibration}
\label{fig:CPFE_Calibration}
\end{figure}

\begin{figure}[htbp]
\centering
\includegraphics[width=0.9\textwidth]{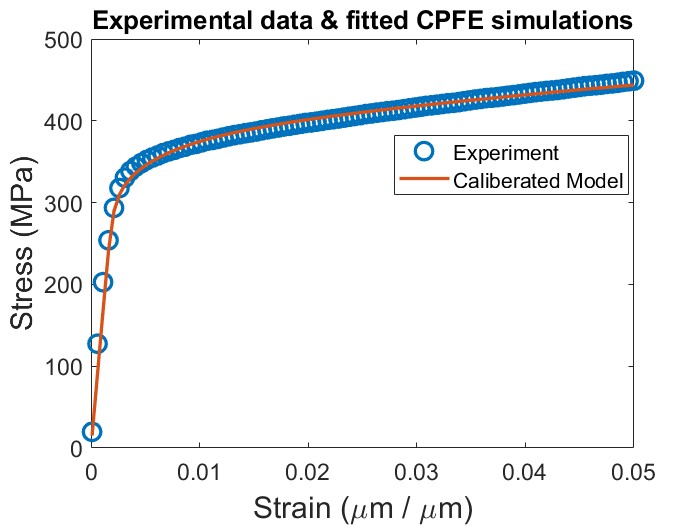}
\caption{Calibrated CPFE response and experimental data}
\label{fig:CPFE_Calibration1}
\end{figure}

\subsection{RVE Size and Effective Grains}

The microstructure of a material comprises grains of varying morphologies, sizes, and orientations. The grain size is studied in the section and next section focuses on texture. The grains in the representative volume element (RVE) of a CPFE model are created based on the mean grain diameter and its standard deviation, minimum, and maximum cutoff values of the distribution. What this means is that, due to the stochastic nature of the input for grain size, the results of the simulation exhibit variability. In the context of degradation modeling, this variability in stress and strain responses can strongly influence predictions of damage initiation and progression. If RVE contains insufficient grains, the variability of the results will be higher and every run with a different set of grains results in a different macromechanical response. Thus, crystal plasticity simulations require sufficient number of grains in the RVE to yield reliable results but not at very heavy computational cost. By increasing the size of the RVE, we can in turn increase the number of grains for the same average grain size. Figure \ref{fig:RVE_SizeGrain}, shows the variability for different RVEs and the number of grains. Consider four different RVEs of 40, 80, 120, and 160 $\mu m$, with approximately 25, 190, 615, and 1400 grains, respectively. For each case, the simulations are run 100 times, and the figure presents the mean and corresponding standard deviation for all of these cases. As observed, the variability decreases as the number of grains increases. This convergence is particularly important when quantifying uncertainty in degradation-sensitive metrics such as yield stress and local strain accumulation. Furthermore, Figures \ref{fig:RVE_SizeGrain1} and \ref{fig:RVE_SizeGrain2} display the stress values at 0.2\% (yield stress), 1.0\%, and 3.0\%, along with their associated variability and corresponding error. Interestingly, there is no significant reduction in variability when comparing the 120 $\mu m$ RVE with the 160 $\mu m$ RVE. This suggests that beyond a certain grain count, the added computational cost does not meaningfully enhance degradation-relevant accuracy. However, it is pertinent to note that the computation time for the 160 $\mu m$ RVE simulation is around four times higher than that for the 120 $\mu m$ RVE. Considering the trade-off between computational cost and accuracy, we can conclude that the RVE containing approximately 615 grains, with a size of 120 $\mu m$, is well-suited for future analyses. This RVE size ensures reliable prediction of degradation precursors, while remaining computationally feasible for uncertainty quantification studies. A compelling visual representation of the final RVE of 615 grains is captured in Figure \ref{fig:RVE_615}.

\begin{figure}[htbp]
\centering
\includegraphics[width=0.9\textwidth]{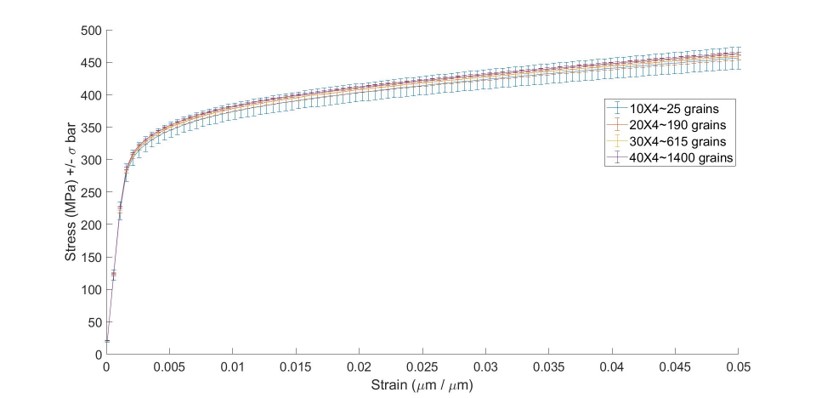}
\caption{Different RVE with sizes of 40 $\mu$m, 80 $\mu$m, 120 $\mu$m and 160 $\mu$m  (100 realizations for each case).}
\label{fig:RVE_SizeGrain}
\end{figure}

\begin{figure}[htbp]
\centering
\includegraphics[width=0.9\textwidth]{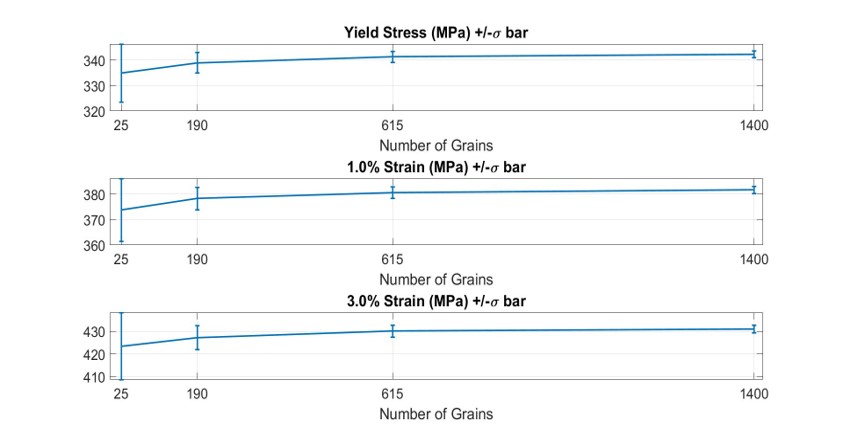}
\caption{CPFE response with number of grains (and RVE size): stress with number of Grains}
\label{fig:RVE_SizeGrain1}
\end{figure}

\begin{figure}[htbp]
\centering
\includegraphics[width=0.9\textwidth]{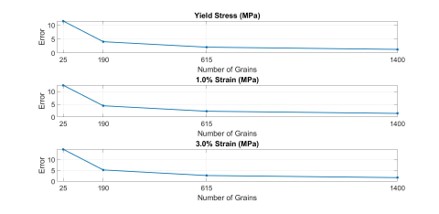}
\caption{CPFE response error with number of grains (and RVE size): associated scattering}
\label{fig:RVE_SizeGrain2}
\end{figure}

\begin{figure}[htbp]
\centering
\includegraphics[width=0.9\textwidth]{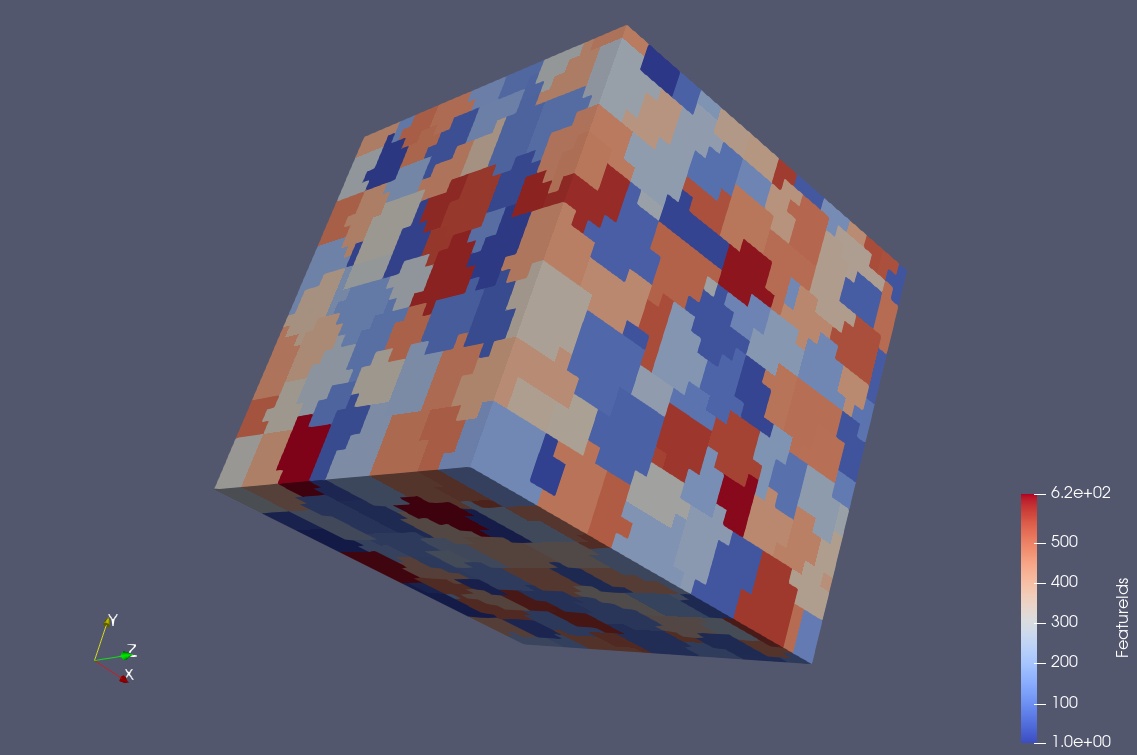}
\caption{RVE with 615 grains considered for the analysis}
\label{fig:RVE_615}
\end{figure}

\subsection{Texture and Grains Orientation Angle}

The properties of a material on a macroscopic scale are intricately linked to the orientations of its individual grains. These grains tend to cluster in specific orientations known as texture components, such as Brass, Goss, Copper, and Cube. For example, in the brass texture, the orientation angles (or Euler angles) in three dimensions are distributed around (30°, 45°, 0°), while in Copper they are around (90°, 35°, 45°), and in Cube they are around (0°, 0°, 0°). Euler angles are vital for texture analysis in materials science, representing the orientation of a crystal in a polycrystalline material. They consist of three angles: $\theta_1, \theta_2$, and $\theta_3$, corresponding to sequential rotations. These angles precisely describe the orientation of each grain, allowing researchers to analyze the texture of the material, which influences anisotropy in plastic response and the initiation of degradation mechanisms such as strain localization and microcrack formation. This data is essential for crystal plasticity finite element modeling to predict material behavior under stress. In degradation-sensitive applications, accurately capturing grain orientation distributions is crucial for understanding variability in failure modes. To understand these texture components in depth, 3D orientation angle data for various texture components are gathered from representative volume elements (RVE) containing 28 grains, and these texture components are visually represented in Figure \ref{fig:Texture}. For that, 200 samples of different texture orientations angles in 3D for Brass (case 1), Copper (case 2), Cube (case 3), and without preference (case 4) in an RVE with 28 grains are extracted from the dream3D tool.

To study the influence of these texture components on elastic and plastic material behavior, strain stress data are extracted from CPFE simulations for each case. This involves obtaining stress-strain data from 200 samples for each texture orientation and calculating the mean stress and standard deviation. The stress-strain curves and scatterings for each preferred orientation are analysed and presented in Figures \ref{fig:SS1}, \ref{fig:SS2}, and \ref{fig:SS3}. Furthermore, the mean stress-strain behavior for these different cases is depicted in Figure \ref{fig:SS4}, providing a detailed comparison of stress-strain curves in different texture orientations. The variations observed in stress responses across different texture clusters highlight the strong role of crystallographic orientation in modulating degradation-relevant responses such as yield stress variability and early plastic strain distribution. In summary, these analyses underscore the impact of grain orientations on the overall mechanical properties of material, with direct implications for predicting degradation susceptibility in textured microstructures.

\begin{figure}[htbp]
\centering
\includegraphics[width=0.5\textwidth]{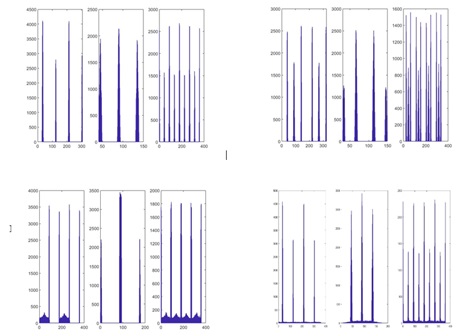}
\caption{200 samples of different orientation, Brass: [$35^{\circ}$,$45^{\circ}$,$0^{\circ}$] on top left, Copper: [$90^{\circ}$,$35^{\circ}$,$45^{\circ}$] on top right, Cube: [$0^{\circ}$,$0^{\circ}$,$0^{\circ}$] on bottom left, and no preference on bottom right}
\label{fig:Texture}
\end{figure}

\begin{figure}[htbp]
\centering
\includegraphics[width=0.5\textwidth]{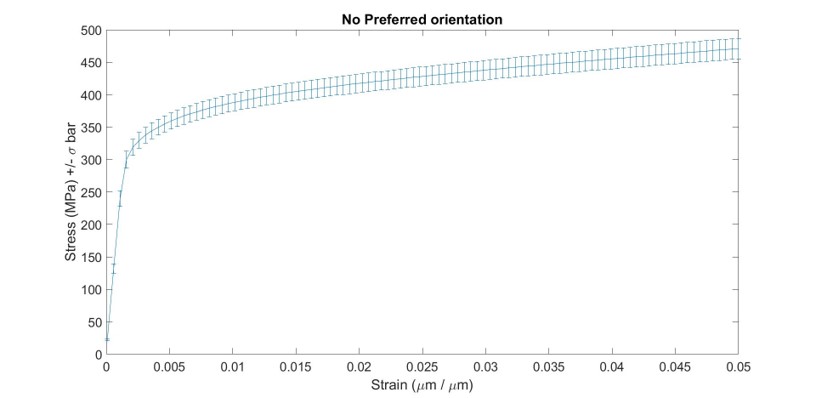}
\caption{Stress-strain curve with no preferred grains orientation and associated scattering}
\label{fig:SS1}
\end{figure}

\begin{figure}[htbp]
\centering
\includegraphics[width=0.5\textwidth]{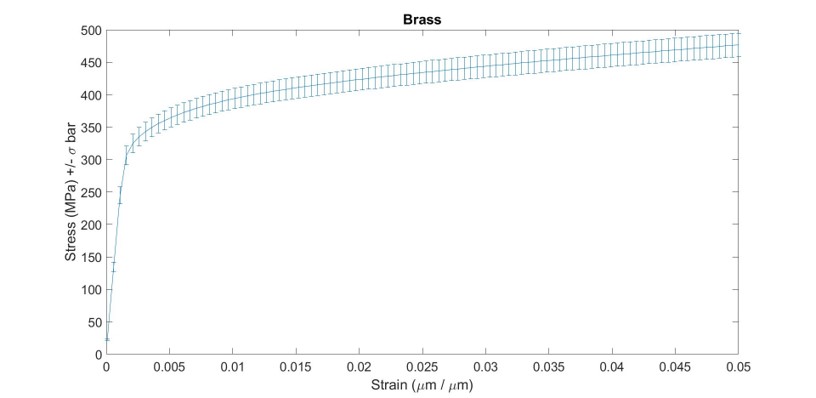}
\caption{Stress-strain curve with brass grains orientation and associated scattering}
\label{fig:SS2}
\end{figure}

\begin{figure}[htbp]
\centering
\includegraphics[width=0.5\textwidth]{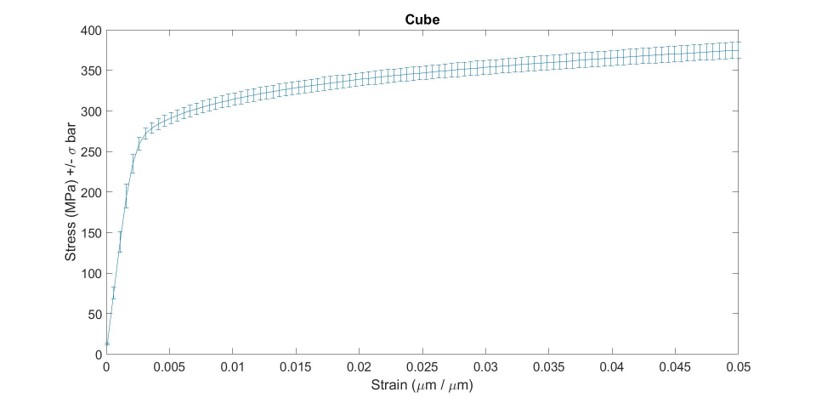}
\caption{Stress-strain curve with cube grains orientation and associated scattering}
\label{fig:SS3}
\end{figure}

\begin{figure}[htbp]
\centering
\includegraphics[width=0.5\textwidth]{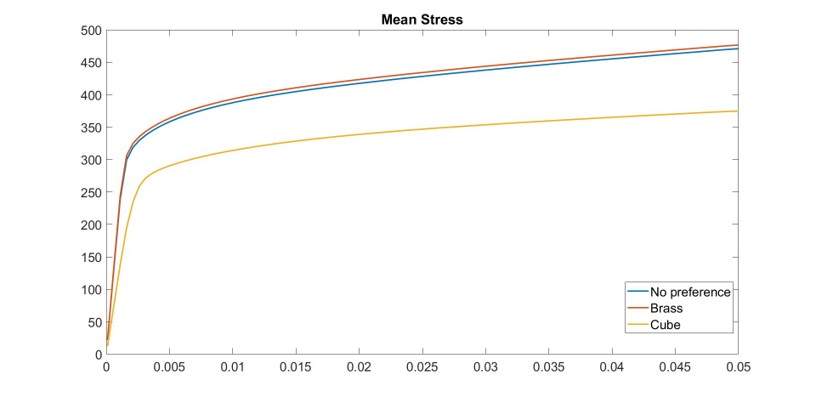}
\caption{Mean stress-strain curve for different grains orientations}
\label{fig:SS4}
\end{figure}

\subsection{Texture Distribution}

To study the influence of texture on material properties, our initial step is to assess the range of texture in service components as not all theoretical textures are practically and physically possible. Understanding the realistic texture bounds is particularly critical in degradation studies, where certain orientations may amplify strain localization or fatigue sensitivity. This information can be gathered from the literature, expert insights, or experimental data. In this particular study, we utilized Electron Back-Scatter Diffraction (EBSD) data from the Stainless Steel type 316L electron beam welded specimen to estimate the texture bounds (see section Experiments). This process involved a detailed analysis and calculation, as illustrated in Figure \ref{fig:EBSDMaps}, which presents an EBSD sample for the SS316 specimen.

We collected a total of thirteen EBSD maps, with each map divided into eight sub-sections. This approach ensured that we considered 104 cases to capture the range of texture components in the weldment. This wide sampling helps account for spatial heterogeneity in the weld microstructure, which directly impacts localized mechanical degradation. The texture information for each case was extracted using the open source software ATEX, and the results are presented in Figure \ref{fig:EBSD1}, showing the percentage of different texture components for a sample section taken from the EBSD map, as shown in Figure \ref{fig:EBSDMaps}. In particular, we identified eight types of texture components (Cb, Gs, Bs, S1, S2, S3, Cu, Ty) in the sample.

Following a similar process, we systematically processed all 104 cases to extract the volume fraction of the texture components. These values were then visualized in a histogram, as depicted in Figure \ref{fig:TextureDistribution}, which delineates the minimum and maximum possible values of different texture components and their distributions in the sample. These variations define the envelope of texture-induced mechanical behavior variability and serve as a degradation-relevant uncertainty input. These variations in texture components will serve as a fundamental input for building a surrogate model. Furthermore, we tabulated the extracted bands of the texture components in Table \ref{tab:TextureInputs}, which shows the minimum and maximum volume fractions of all the texture components in the sample. This data provides physically grounded constraints that enable more realistic degradation modeling through uncertainty-aware simulations.

\begin{figure}[htbp]
\centering
\includegraphics[width=0.8\textwidth]{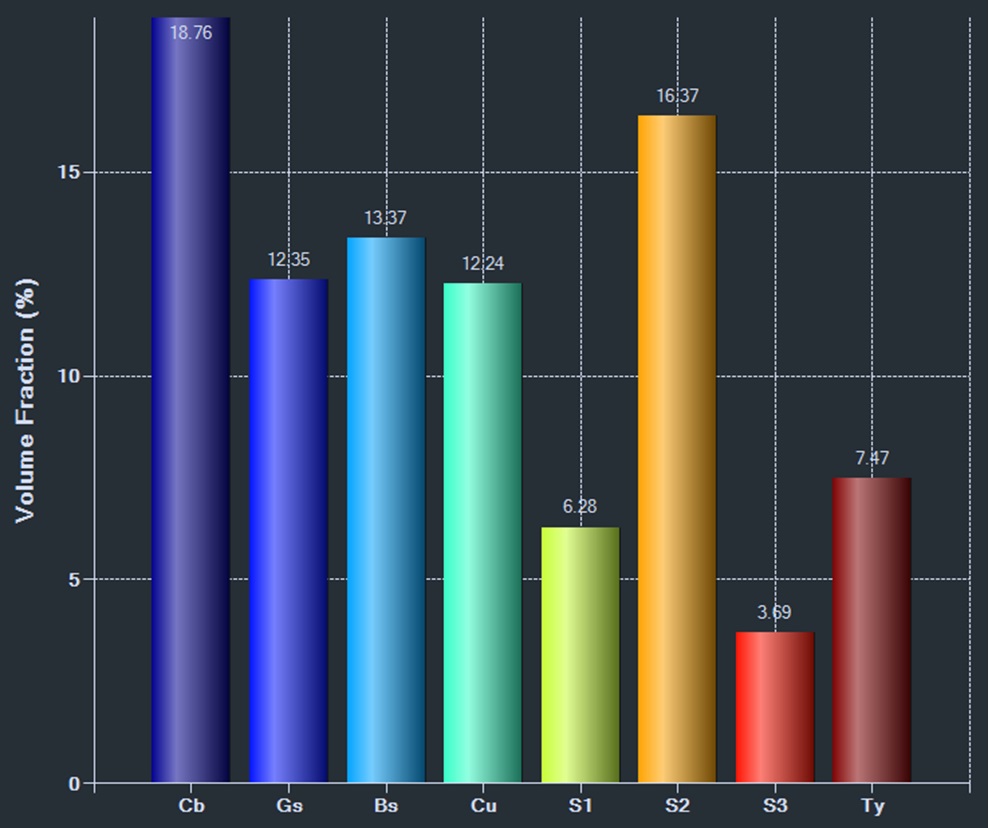}
\caption{Texture information for a sample taken from a section as shown in Figure \ref{fig:EBSDMaps}}
\label{fig:EBSD1}
\end{figure}

\begin{figure}[htbp]
\centering
\includegraphics[width=0.9\textwidth]{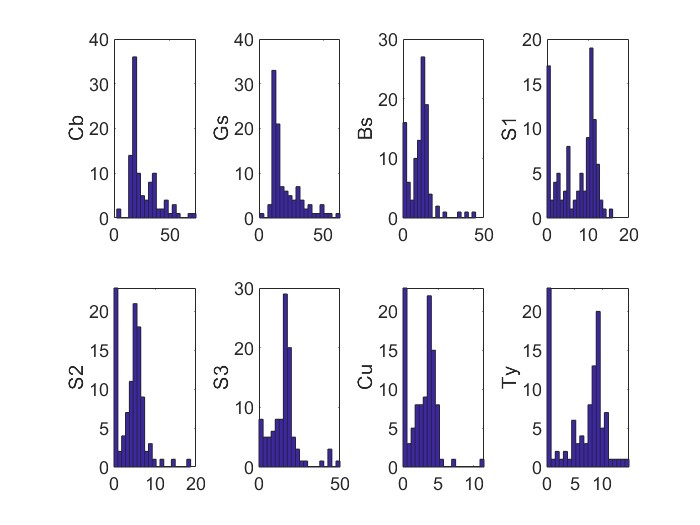}
\caption{Texture distribution (histogram) extracted from EBSD data}
\label{fig:TextureDistribution}
\end{figure}

\begin{table}[!ht]
\centering
\caption{Texture bounds extracted from EBSD samples }
\begin{tabular}{lllllllll}
 \hline\\
Texture & Cb & Gs & Bs & S1 & S2 & S3 & Cu & Ty\\
     \hline\\
min & 2.5 & 0.71 & 0 & 0.01 & 0 & 0.02 & 0 & 0\\
max & 73.81 & 63.25 & 44.86 & 16.01 & 18.68 & 50.25 & 11.48 & 14.86\\
\hline\\
\end{tabular}
\label{tab:TextureInputs}
\end{table}

\subsection{Surrogate Model and Validation}

In the previous section, we discussed the presence of eight different texture components (Cube, Goss, Brass, S1, S2, S3, Copper and Taylor) in the SS316 sample. To construct a surrogate model and perform texture analyses, we considered eight input parameters and utilized the open source Matlab toolbox, MTex, to create a joint probability distribution of the texture components. These input parameters were varied on the basis of the minimum and maximum texture values, and MTEX was employed to calculate the corresponding Euler angles for a given set of weights. Subsequently, 200 input samples (sets of weights) were randomly generated, and the corresponding Euler angles were calculated to generate 200 Representative Volume Elements (RVE) with different texture components. These diverse RVEs reflect the realistic texture heterogeneity observed in welded components, which is critical for understanding how texture variability contributes to degradation susceptibility. We conducted simulations using the crystal plasticity model, which is a UMAT subroutine of Abaqus, for different texture inputs to obtain the stress-strain data for all 200 simulations. The input weights and corresponding stress-strain data were then utilized to construct a polynomial chaos-based surrogate model as described previously. This surrogate is designed to efficiently capture how microtexture variations drive uncertainty in degradation-relevant responses such as yield stress and strain localization. This model was trained with 80\% randomly selected samples, with the remaining 20\% samples used for validation. Validation, as shown in Figure \ref {fig:SurrogateR}, involved displaying stress-strain curves for eight randomly selected samples (sample numbers 29, 35, 2, 9, 17, 34, 27, 15) out of 40 samples unknown to the surrogate model, along with the stress-strain data predicted by the surrogate model for these samples. The close agreement between simulated and predicted curves confirms that the surrogate model reliably approximates texture-induced variability in mechanical behavior—essential for degradation-aware simulations.

\begin{figure}[htbp]
\centering
\includegraphics[width=0.9\textwidth]{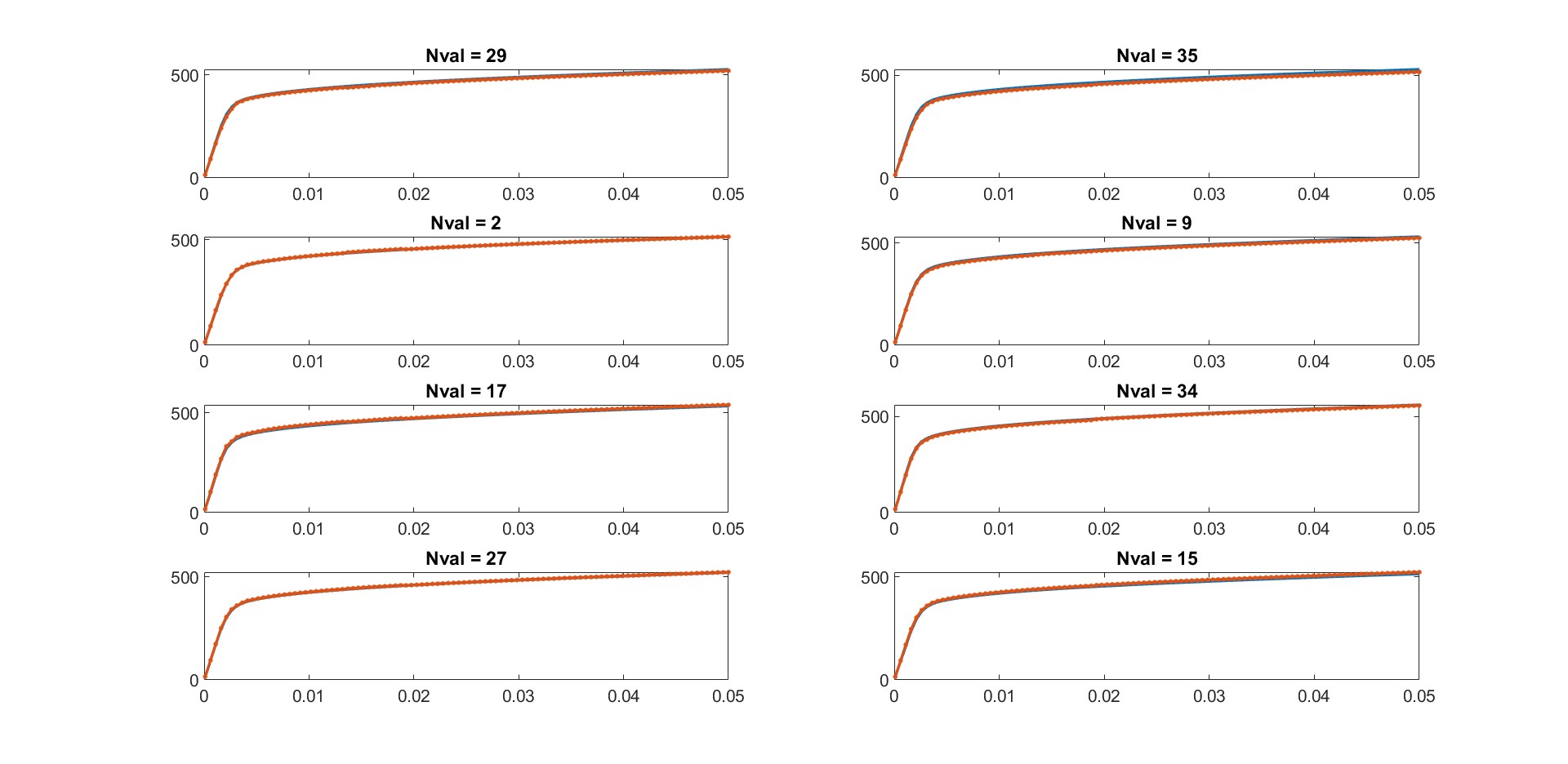}
\caption{Results from the surrogate model against randomly selected 8 validation samples}
\label{fig:SurrogateR}
\end{figure}

\subsection{Impact of Texture on Stress-strain}

Our investigation of the influence of texture on stress-strain response, using the surrogate model developed in the preceding section, has revealed significant findings. The model's response to texture is assessed in terms of mean stress concerning strain and uncertainty bounds in stress. This is particularly important in the context of degradation modeling, where localized stress variations can initiate damage mechanisms such as microcrack nucleation or fatigue. The yield stress and stresses at two other strain points are also examined for both mean values and their corresponding uncertainty bounds. In Figure \ref{fig:SurrogateR1}, the mean strain curve is presented, along with its uncertainty limits. Estimation of the mean and standard deviation for stress data is performed using polynomial chaos theory, as elaborated in the previous section. In Figure \ref{fig:SurrogateR11}, stress values and their uncertainty bounds are estimated at three different strain points (0.2\%, 1.0\%, and 3.0\%). The corresponding values are also tabulated in Table \ref{tab:StressValues}. To identify the stress points, lines parallel to the slope of Young's modulus are drawn at stress levels of 0.2\%, 1.0\%, and 3.0\%. The growing uncertainty with increasing strain indicates rising variability in plastic deformation, which directly correlates with early degradation susceptibility under service loading. It is noticeable that the uncertainties are smaller for the yield stress at the 0.2\% strain point and larger at the 3.0\% strain point. Furthermore, the sensitivity of the texture components to stress is illustrated for each point of strain. It is evident from the analysis that the Cube and Goss texture components exert the most significant influence. These components strongly affect the stress field evolution and are likely contributors to localized degradation behavior, such as anisotropic fatigue or creep damage. That means that for a given material specimen, small alterations in either the Cube or Goss texture can lead to substantial changes in material properties. In contrast, the Brass and S1 texture components exhibit almost negligible sensitivity, indicating that variations in these texture components have negligible impact on the properties of the material. The remaining components of the texture (S2, S3, Copper, and Ty) demonstrate minimal effects on stress. This suggests that not all texture variability contributes equally to degradation risk, and prioritizing critical orientations can improve predictive accuracy. This implies that two different SS316 samples with a small difference in their Cube and Goss texture components will make a big difference in their material properties. However, if two material samples share the same Cube and Goss texture but differ in the components of the Brass, S1, S2, S3, Copper, and Taylor texture components, their material properties will be nearly identical. In summary, our research has shown that the Cube and Goss texture components are the most sensitive, whereas other texture components have almost no impact on material properties. These insights are essential for guiding microstructure-sensitive degradation assessments and tailoring weld procedures to control texture-driven failure variability.

\begin{figure}[htbp]
\centering
\includegraphics[width=0.9\textwidth]{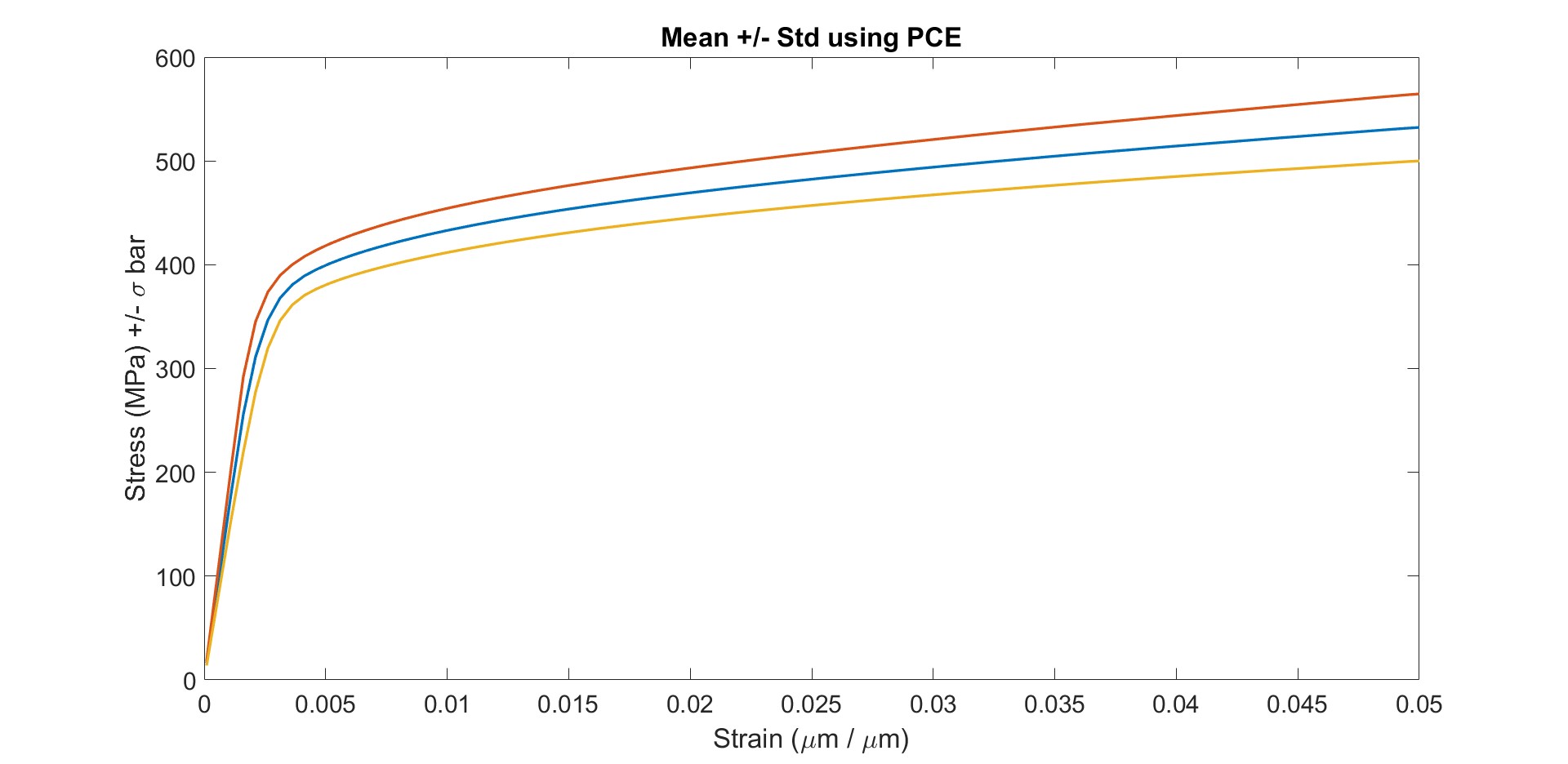}
\includegraphics[width=0.9\textwidth]{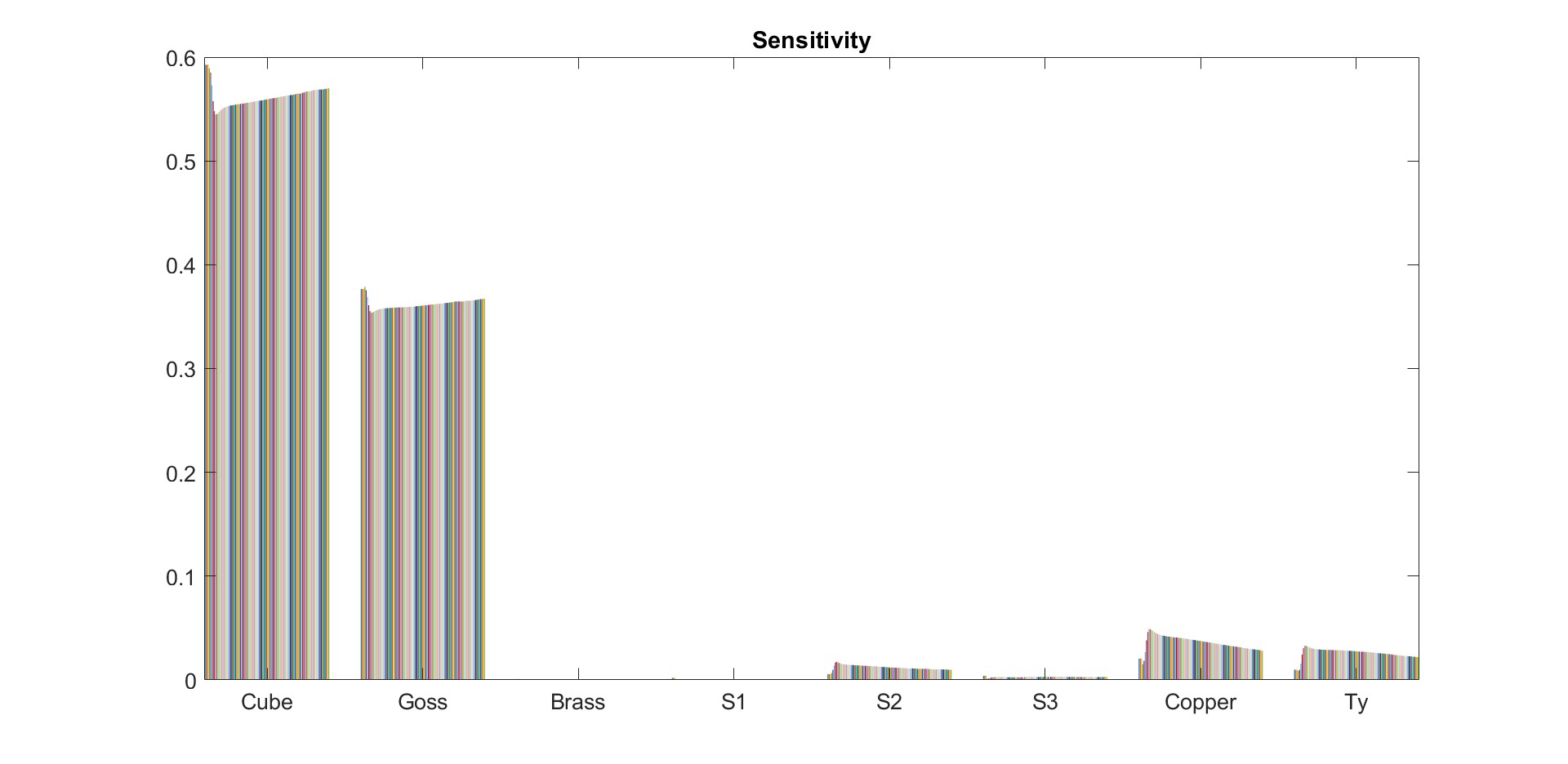}
\caption{Statistical analysis and impact of texture on the stress-strain: mean and variability (top) and sensitivity analysis (bottom)}
\label{fig:SurrogateR1}
\end{figure}

\begin{figure}[htbp]
\centering
\includegraphics[width=0.9\textwidth]{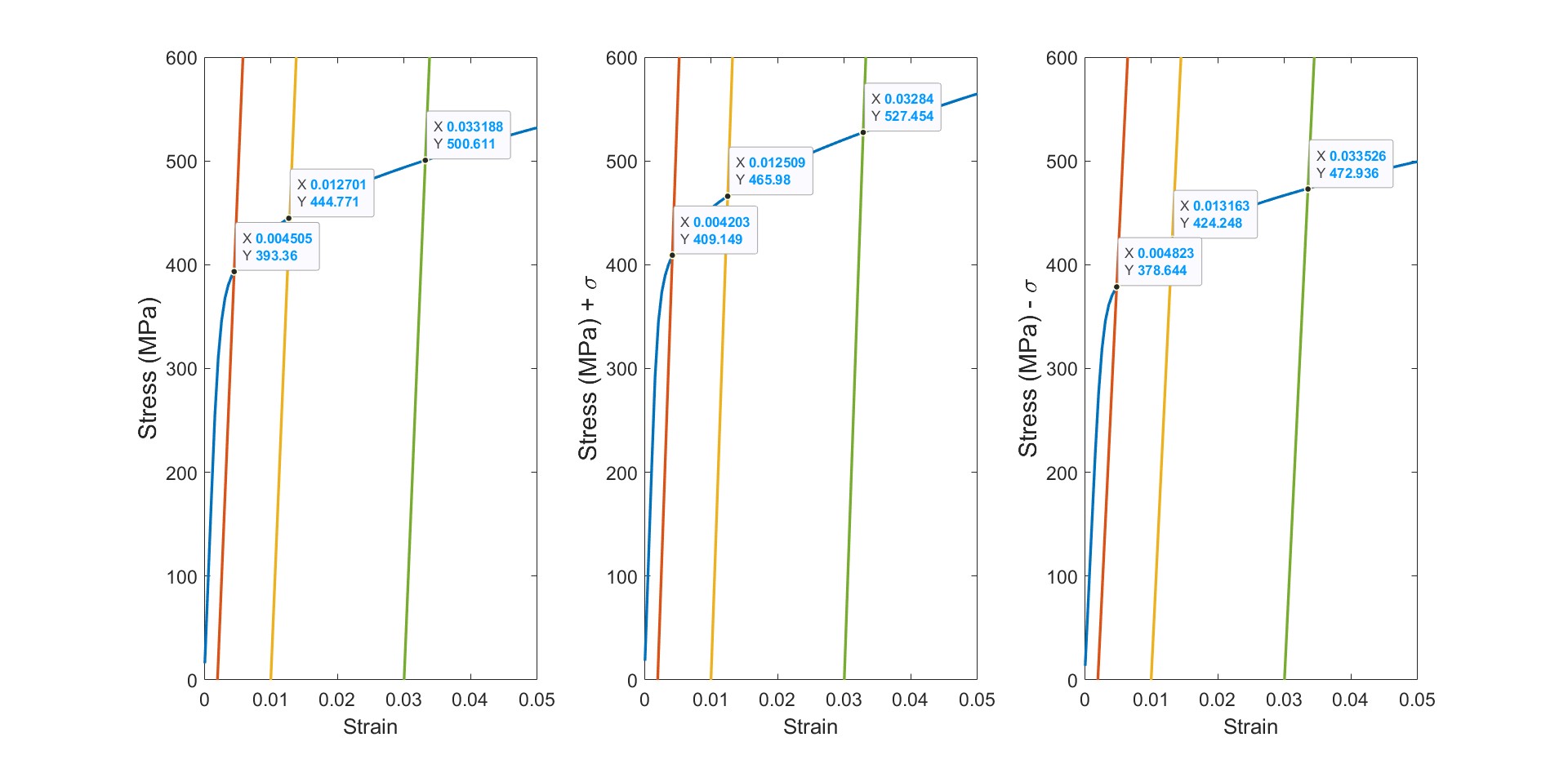}
\caption{Stress values at 0.2\% (Yield stress), 1.0\% and 3.0\% strain}
\label{fig:SurrogateR11}
\end{figure}

\begin{table}[!ht]
\centering
\caption{Stress values and their bound of uncertainties at 0.2\% (Yield strain), 1.0\% and 3.0\% strain}
\begin{tabular}{lllllll}
 \hline\\
Stress Values & Mean (in MPa) & Min. (in MPa) & Max. (in MPa)\\
     \hline\\
Yield stress& 393.36 & 378.64 & 409.15 \\
At 1.0\% strain& 444.77 & 424.25 & 465.98 \\
At 3.0\% strain& 500.61 & 472.94 & 527.45 \\
\hline\\
\end{tabular}
\label{tab:StressValues}
\end{table}

\section{Conclusions}

This study presents a first-of-its-kind, degradation-aware framework that integrates crystal plasticity finite element (CPFE) modeling with a physics-informed polynomial chaos surrogate to systematically quantify how crystallographic texture influences early-stage degradation behavior in 316L stainless steel weldments. After a rigorous mesh refinement study, a representative volume element (RVE) comprising 615 grains was selected to enable statistically converged predictions under texture variability. The CPFE model was meticulously calibrated against experimental stress-strain data, and a sparse polynomial chaos-based surrogate was trained using only 200 simulations—offering orders-of-magnitude speedup relative to traditional Monte Carlo-based uncertainty quantification.

Critically, the surrogate model preserved the fidelity of CPFE outputs while enabling rapid stochastic analyses of texture-induced plastic response, capturing both mean stress behavior and variability bounds that are directly linked to fatigue, strain localization, and crack initiation risks. Texture bounds used for input were extracted from high-throughput EBSD maps of electron beam welded samples, ensuring physical realism in the surrogate's uncertainty representation.

Sensitivity analysis unequivocally identified Cube and Goss texture components as dominant drivers of stress variability, highlighting their central role in microstructure-sensitive degradation pathways. In contrast, components such as Brass, S1, S2, S3, Copper, and Taylor exhibited minimal impact, offering a clear hierarchy for microstructural control in welding and post-processing.

This integrated methodology advances the predictive accuracy, computational tractability, and physical interpretability of degradation simulations in nuclear-grade structural alloys. It provides a scalable platform for assessing risk, optimizing weld parameters, and designing microstructures with superior degradation resistance, making it highly relevant for high-reliability sectors such as nuclear energy, aerospace, and critical infrastructure.

\section*{Acknowledgments}

 The authors appreciate the support of UKRI through SINDRI (EP/V038079/1) and financial support from EDF Energy. MM Acknowledges the support of Royal Academy of Engineering through a Research Chair scheme. Henry Royce Institute is gratefully acknowledged for the provision of advanced microscopy facilities at The University of Manchester, funded through EPSRC grants EP/R00661X/1, EP/S019367/1, and EP/P025498/1. Nuclear AMRC is gratefully acknowledged for manufacturing the weldements under Project FORCE. FORCE  was commissioned by the UK Government's Department of Business, Energy and Industrial Strategy Nuclear Innovation Programme. The views expressed in the paper are those of the authors and should not be interpreted as BEIS or wider Government policy. FORCE was conducted and managed by Frazer-Nash Consultancy who are gratefully acknowledged.

\section*{Data and code availability}
The data and code used and/or analyzed during this study are available from the corresponding author on reasonable request.

\section*{Competing Interests}
The authors declare no conflict of interest.


\section*{Declaration of Generative AI and AI-assisted technologies in the writing process}
During the preparation of this work the author(s) used ChatGPT in order to language editing and refinement. After using this tool/service, the author(s) reviewed and edited the content as needed and take(s) full responsibility for the content of the publication.

\bibliographystyle{unsrtnat}
\bibliography{references} 

\end{document}